\documentclass[twocolumn]{aastex631}
\usepackage{apjfonts}
\usepackage{graphics,graphicx,float,import,enumitem,booktabs}

\newcommand\hii{\ion{H}{2} }
\newcommand\heii{\ion{He}{2}}
\newcommand\oii{[\ion{O}{2}]}
\newcommand\oiii{[\ion{O}{3}]}
\newcommand\oiiisf{\ion{O}{3}]}
\newcommand\ciii{\ion{C}{3}]}
\newcommand\civ{\ion{C}{4}}

\newcommand\nii{[\ion{N}{2}]}
\newcommand\sii{[\ion{S}{2}]}
\newcommand\W{$\lambda$}

\shorttitle{UV Spectroscopy of Pox 186}
\shortauthors{Rogers et al.}
\accepted{to ApJ on August 20th, 2023}

\begin{document}

\title{HST UV Spectroscopy of the Dwarf Starburst Galaxy Pox 186}

\author{Noah S. J. Rogers}
\affiliation{Minnesota Institute for Astrophysics, University of Minnesota, 116 Church St. SE, Minneapolis, MN, 55455, USA}

\author{Claudia M. Scarlata}
\affiliation{Minnesota Institute for Astrophysics, University of Minnesota, 116 Church St. SE, Minneapolis, MN, 55455, USA}

\author{Evan D. Skillman}
\affiliation{Minnesota Institute for Astrophysics, University of Minnesota, 116 Church St. SE, Minneapolis, MN, 55455, USA}

\author{Nathan R. Eggen}
\affiliation{Minnesota Institute for Astrophysics, University of Minnesota, 116 Church St. SE, Minneapolis, MN, 55455, USA}

\author{Anne E. Jaskot}
\affiliation{Department of Astronomy, Williams College, Williamstown, MA, 01267, USA}

\author{Vihang Mehta}
\affiliation{IPAC, Mail Code 314-6, California Institute of Technology, 1200 E. California Blvd., Pasadena, CA, 91125, USA}

\author{John M. Cannon}
\affiliation{Department of Physics \& Astronomy, Macalester College, 1600 Grand Avenue, Saint Paul, MN, 55105, USA }

\begin{abstract}

Studying the galaxies responsible for reionization is often conducted through local reionization-era analogs; however, many of these local analogs are too massive to be representative of the low-mass star-forming galaxies that are thought to play a dominant role in reionization. The local, low-mass dwarf starburst galaxy Pox 186 is one such system with physical conditions representative of a reionization-era starburst galaxy. We present deep ultraviolet (UV) spectroscopy of Pox 186 to study its stellar population and ionization conditions and to compare these conditions to other local starburst galaxies. The new Cosmic Origins Spectrograph data are combined with archival observations to cover $\sim$1150–2000 \AA\ and allow for an assessment of Pox 186's stellar population, the relative enrichment of C and O, and the escape of ionizing photons. We detect significant Ly$\alpha$ and low-ionization state absorption features, indicative of previously undetected neutral gas in Pox 186. The C/O relative abundance, log(C/O) $=$ $-$0.62$\pm$0.02, is consistent with other low-metallicity dwarf galaxies and suggests a comparable star formation history in these systems. We compare UV line ratios in Pox 186 to those of dwarf galaxies and photoionization models, and we find excellent agreement for the ratios utilizing the intense \ciii, \oiiisf, and double-peaked \civ\ lines. However, the UV and optical \heii\ emission is faint and distinguishes Pox 186 from other local starburst dwarf galaxies. We explore mechanisms that could produce faint \heii, which have implications for the low-mass reionization-era galaxies which may have similar ionization conditions.

\end{abstract}

\section{Introduction}

Reionization represents the last great phase change of the universe, where the H$^0$-dominated universe transitions back to being predominantly ionized. One contributor to reionization is the production of ultraviolet (UV) photons from early star-forming galaxies. These galaxies, with low stellar mass and significant escape fractions of ionizing radiation, are thought to be a primary mechanism for the reionization of the universe \citep[e.g.,][]{live2017,weis2017,stei2018}. However, the study of these high-$z$ galaxies is hampered by intervening neutral gas, and broad wavelength coverage is necessary to observe the emission lines that reliably trace gas-phase properties. With the successful launch of the \textit{James Webb Space Telescope}, it is now possible to observe some of these extreme galaxies and constrain the gas-phase physics and chemical enrichment in reionization-era systems \citep[e.g.,][]{arel2022,scha2022}. However, low-$z$ analogs to reionization-era galaxies present the best opportunity to reliably measure these properties in a significant number of sources and on small spatial scales that are inaccessible at high-$z$. To understand the conditions that led to reionization, it is vital to obtain panchromatic information on local reionization-era analogs.

Perhaps the most notable sample of reionization-era analogs are the Green Pea (GP) galaxies, so named for their extreme emission from the optical \oiii\ lines and compact structure \citep{card2009}. Other properties that characterize these galaxies are highly-ionized gas (as determined via $O_{32}$ = \oiii$\lambda$5007/\oii$\lambda$3727) and low oxygen abundances \citep[e.g.,][]{izot2011,hawl2012,jask2019}. The original sample of GPs was comprised of galaxies at 0.1$\lesssim z\lesssim$0.35, a redshift range where the strong \oiii\ lines are redshifted into the Sloan Digital Sky Survey (SDSS) $r$-band; galaxies at z$\gtrsim$0.3 are particularly useful for studying the escape of ionizing radiation, as it is possible to directly measure their Lyman Continuum (LyC) photons with instruments on the \textit{Hubble Space Telescope} (HST). Indeed, the GPs and other compact, highly star-forming (SF) galaxies have been observed to have relatively high escape fractions between 2 and 79\% \citep{izot2018a,izot2018b}. Furthermore, nine of the original GPs have nebular emission from He$^{2+}$ \citep{hawl2012}, implying the significant production of photons with energies $>$ 54.4 eV. \citet{jask2013} explore the high-ionization emission in the six GPs with the largest $O_{32}$ ratios, finding that five of these extreme GPs show significant optical \heii\ emission. Modeling these spectra requires a young stellar population of 3-5 Myr, and a single burst of star formation can result in the intense \heii\ emission if the optical depth is high or if the filling factor is less than unity \citep{jask2013}. Many of these conditions are expected in the galaxies responsible for reionization, but the GPs have masses that are significantly larger than expected for these early galaxies. Low-mass objects with similar conditions at high-$z$ are challenging to observe; while still rare, low-mass local analogs to reionization-era galaxies provide the same utility as the GPs and permit a thorough investigation of galaxy evolution at low mass.

One extreme example of a local, low-mass galaxy with properties similar to those expected for reionization era systems is the dwarf starburst galaxy Pox 186. \citet{kunth1981} first noted this galaxy's compact nature and intense \oiii\ emission, but Pox 186 has been undetected in \ion{H}{1} 21cm observations despite numerous attempts \citep{kunth1988,begu2005}. While recent MeerKAT \ion{H}{1} 21cm observations indicate a detection of neutral gas (Cannon et al. in preparation), the low \ion{H}{1} mass might suggest a density-bounded ionization structure which could permit a large escape of ionizing radiation. Following these early works, \citet{corb2002} obtained HST Planetary Camera 2 photometry and Space Telescope Imaging Spectrograph (STIS) UV spectroscopy of Pox 186, and complemented these data with optical spectroscopy from the Bok 90" telescope. The optical spectroscopy and photometry indicate a low-metallicity, highly-ionized environment that is ionized by a young (4$^{+1}_{-1.6}$ Myr) stellar population with a low total stellar mass of log(M$_*$/M$_\odot$) $\sim$ 5. While \citet{corb2002} report tentative detections of Ly$\alpha$ and \ciii$\lambda\lambda$1906,1909 emission, the STIS spectrum is too noisy and at insufficient resolution to resolve and robustly fit these emission lines.

\citet{guse2004} obtained additional optical spectroscopy of Pox 186 using the Multiple Mirror Telescope's (MMT's) blue channel spectrograph (3650 - 7500 \AA, spectral resolution $\sim$7 \AA) and the 3.6m ESO telescope's grism \#11 (3400 - 7400 \AA, spectral resolution $\sim$13 \AA). These spectra extend to lower wavelengths and include the \oii$\lambda\lambda$3726,3729 doublet necessary for accurate O/H abundance determinations. \citet{guse2004} found that the gas-phase oxygen abundance of Pox 186 is 12+log(O/H) = 7.74$-$7.76 dex, $O_{32}$ $>$ 18, the S/O abundance is enhanced, and the spectra contain emission from high ionization species such as Ne$^{2+}$, Ar$^{3+}$, and He$^{2+}$. The large $O_{32}$ and enhanced S/O abundance are both expected in a density-bounded environment where the amount of gas containing the low-ionization ions is less than the high-ionization gas, producing low \oii\ emission and resulting in an erroneous S Ionization Correction Factor (ICF). Furthermore, the detection of significant, nebular \heii$\lambda$4686 is only possible if there is production of very high energy photons from a hard ionizing continuum. Finally, \citet{egge2021} presented recent Gemini Multi-Object Spectrograph (GMOS) Integral Field Unit (IFU) spectroscopy of Pox 186 and uncovered a significant outflow component in the warm gas, another potential mechanism for the escape of ionizing radiation and further supporting the claim that Pox 186 is a local reionization-era analog.

While the optical and radio observations have provided hints concerning the chemical abundances and ionizing structure in the Interstellar Medium (ISM) of Pox 186, there is much to be gained from UV observations of this unique object. First, the FUV contains Ly$\alpha$ at 1216 \AA\ which provides another probe of the escape of ionizing radiation. Prior studies of other reionization-era analogs have observed double-peaked Ly$\alpha$ indicative of the escape of resonantly scattered Ly$\alpha$ emission and low column densities of neutral H \citep{izot2016a,izot2016b,jask2019}. The Ly$\alpha$ profile and the presence of absorption lines from low-ionization metal ions reveal aspects of the neutral medium through which LyC photons would have to escape to ionize the surrounding gas.

Second, emission lines from multiple ionization states of C are observable in the UV and permit a calculation of the  C/O relative abundance. A measure of the C abundance from an optical spectrum is only possible via the detection of \ion{C}{2} recombination lines; these lines are particularly faint (of order 0.1\% the intensity of H$\beta$) and are difficult to observe in low-metallicity objects due to the linear dependence on metal abundance. The semi-forbidden lines \ciii$\lambda\lambda$1906,1909 are intense NUV emission lines that provide an electron density estimate and the C$^{2+}$/O$^{2+}$ relative abundance when compared to the strong \oiiisf$\lambda\lambda$1660,1666 lines. The C/O abundance is dependent on aspects of the star formation history of a galaxy, including the number of bursts, burst timescale, and metal retention in the ISM \citep[see discussion in][]{berg2019}. Constraining this relative abundance can reveal if the star formation history and C enrichment in Pox 186 is consistent with other low-metallicity, highly-SF galaxies. Additionally, the UV \civ\ lines can assess the average ionization within the galaxy and the escape of resonantly scattered photons from the high-ionization gas.

Finally, the expected intensity of the \heii$\lambda$1640 emission line is many times larger than that of the optical \heii$\lambda$4686 line. This emission line can both constrain the observed optical \heii\ emission and can distinguish between SF and active galactic nuclei (AGN) ionizing sources \citep[e.g.,][]{felt2016}. While local Extreme Emission Line Galaxies (EELGs) have significant optical and UV \heii\ emission \citep{berg2021}, it is unclear if the physical conditions in these systems are directly comparable to low-mass reionization-era galaxies. Expanding the sample of UV \heii\ emitters to include low-mass systems permits a critical analysis of the ionizing sources and the production of such high-energy photons in a variety of astrophysical systems.

We present new HST Cosmic Origins Spectrograph (COS) UV spectroscopy of Pox 186 to understand the chemical evolution and ionization conditions in this extreme starburst galaxy and to gain insight into the potential properties of reionization-era systems. In \S2 we describe the COS observations and reductions, the stellar population synthesis modeling to fit the UV continuum, and the emission line fits. \S3 highlights the physical conditions in Pox 186 as measured from the UV emission lines, and we assess the \heii\ intensity measured in the COS spectrum. We compare the emission in Pox 186 to that of other literature SF galaxies and photoionization models, and we discuss the possible production mechanism of faint \heii\ emission in \S4. We summarize our findings in \S5. In this work, we adopt a distance to Pox 186 of 15.3 Mpc \citep{kour2020} and redshift of 0.0041 \citep{egge2021}.

\section{COS Observations and Analysis}

\subsection{UV Spectroscopy}

As part of HST-GO PID 16294 (PI: N. Eggen), we obtained COS G160M observations of Pox 186. These observations cover a wavelength range from 1340 - 1700 \AA\ (central wavelength 1533 \AA). The newly acquired G160M observations are combined with archival G130M and G185M data to cover the rest frame wavelength range between $\sim$1150 and 2000 \AA. The total exposure times  were 5120s for G130M (PI: N. Kumari, PID: 16445), 10460s for G160M (this work), and 13430s for G185M (PI: N. Kumari, PID: 16071). All data were downloaded from the MAST archive and reprocessed using the \textsc{calcos} pipeline v3.4.0 and the most recent calibration files.

\begin{figure}[t]
	\includegraphics[scale=.65]{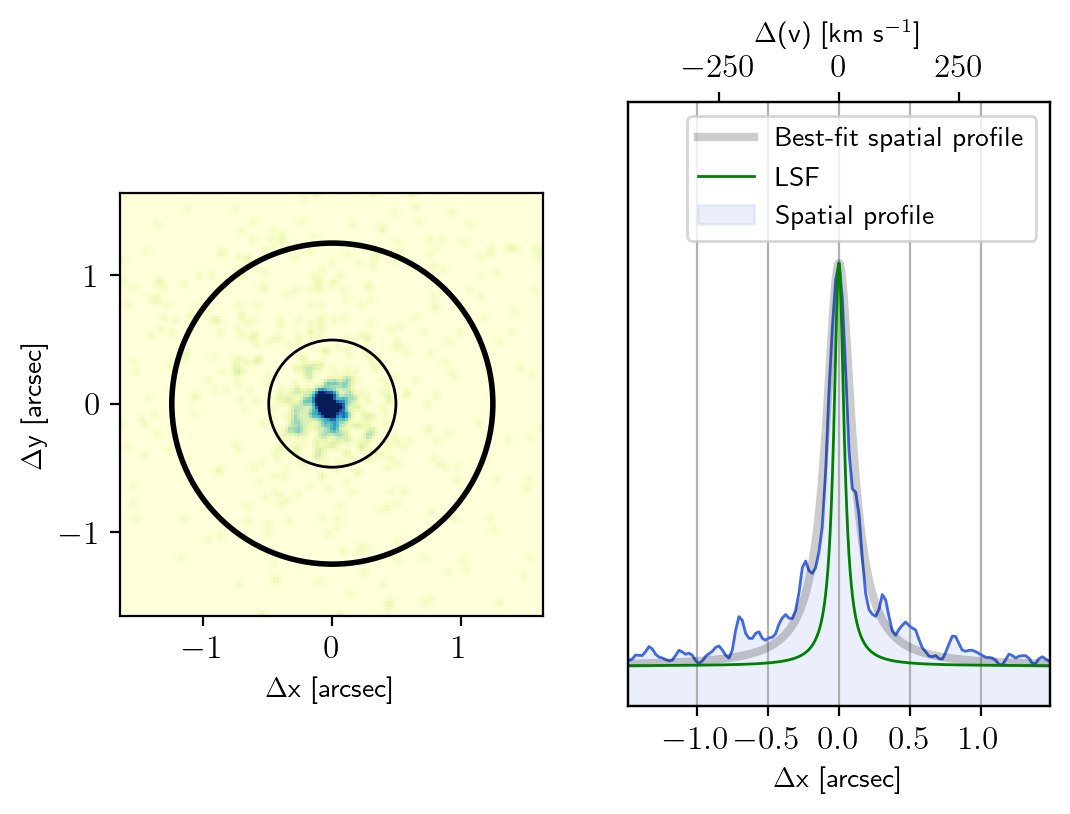}
	\caption{\textit{Left:} NUV acquisition image of Pox 186. \textit{Right:} Normalized light profile in the cross dispersion direction. Pox 186 is not described by a point source (green) but is slightly extended and well-described by a Cauchy distribution (gray).}
	\label{fig:image}
\end{figure}

Figure~\ref{fig:image} shows the NUV acquisition image in the left panel, together with the 1\farcs25 radius COS nominal aperture (thick circle) and the un-vignetted 0\farcs5 radius aperture. Pox 186 is centered in the acquisition image and is clearly compact in the NUV, with $\approx$98\% of the flux included within the un-vignetted aperture. Therefore, we deemed that the default value for the size of the extraction aperture (i.e., a size in the cross dispersion direction of 1\farcs34) was appropriate. We tested using larger extraction apertures and found that the profiles of Ly$\alpha$ and metal lines such as \civ$\lambda\lambda$1548,1551 and \oiiisf$\lambda\lambda$1660,1666 were robust to these changes. Even though the galaxy is compact, it is not a point source, and it is resolved in the COS NUV image. Consequently, the effective spectral resolution of the extracted spectrum is determined by the galaxy's spatial profile in the dispersion direction, rather than the nominal Line Spread Function (LSF) which characterizes the width of spectral features for point sources. In order to estimate the effective spectral resolution, we used the NUV acquisition image and extracted the galaxy's profile summing all the pixels along the cross dispersion direction. The resulting normalized  profile is shown in the right panel of Figure~\ref{fig:image} (blue line).  For comparison, we also show in green the LSF associated with a point source. The galaxy's spatial profile is resolved and well described by a Cauchy distribution with a Full-Width-Half-Maximum (FWHM) of 0\farcs24 \citep[17.8~pc at Pox 186's distance of 15.3 Mpc from][]{kour2020}. The need for using a Cauchy distribution arises because the spatial profile presents broad, low surface brightness,  wings  that cannot be reproduced with a simple Gaussian profile.  In velocity space, the profile's width corresponds to a FWHM of 65 km s$^{-1}$.

\begin{figure*}[t]
   \centering
   \includegraphics[width=0.99\textwidth, trim=40 0 20 0,  clip=yes]{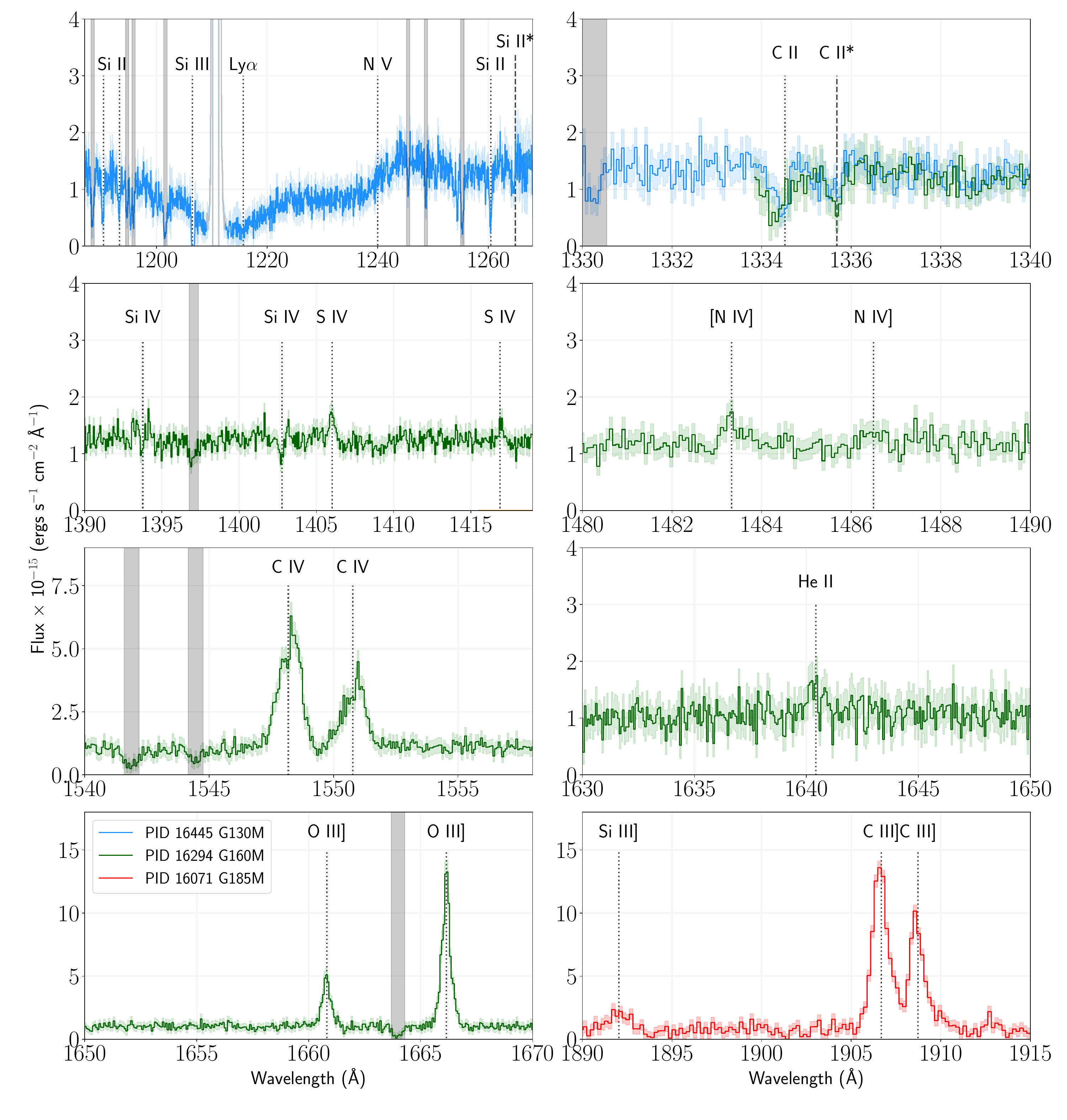}
   \caption{Composite UV spectrum of Pox 186 from multiple HST COS observations. Each panel plots a portion of the composite spectrum corrected for the systemic velocity of Pox 186 and is color coded by the PID and filter of the observation. The G130M and G160M spectra are binned in 6 pixels, the G185M spectrum is binned in 3 pixels. Galactic absorption features are denoted by gray shaded areas and geocoronal Ly$\alpha$ is grayed out. Significant emission/absorption features are shown with dotted lines, fine-structure absorption lines with dashed lines.}
   \label{fig:compspec}
\end{figure*}

We bin the G130M and G160M spectra by 6 COS pixels and the G185M spectra by 3 pixels using the \textsc{Python SpectRes} package \citep{carn2017} which preserves the flux while resampling the flux/error array onto a new wavelength array. Figure \ref{fig:compspec} plots portions of the full UV spectrum, where the different colors represent the different filters/PIDs which obtained the spectra and the shaded regions indicate the uncertainty in the flux measurements. Both the G130M+G160M and G160M+G185M spectra align well at the overlap regions, therefore we do not scale any spectra. The vertical dotted and dashed lines denote key spectral features, while vertical gray shaded areas indicate Galactic absorption lines.

Starting with the G130M spectrum in top left panel, there is Ly$\alpha$ absorption and a broad \ion{N}{5} P Cygni profile at 1240 \AA. These two features are not only blended with each other but are affected by Galactic Ly$\alpha$ absorption and geocoronal Ly$\alpha$ emission. We discuss how these components are fit in \S2.2. Radiative transfer models predict that the line profile of Ly$\alpha$ emission is  double-peaked with a small velocity offset between the peaks when the neutral gas column density is low \citep[e.g.,][]{verh2015}, but Ly$\alpha$ absorption is indicative of neutral gas along the line of sight to absorb the resonantly scattered photons. While Pox 186 has previously been undetected in \ion{H}{1} 21cm observations, the presence of neutral gas would be in agreement with recent MeerKAT detections of \ion{H}{1} 21cm (Cannon et al. in preparation).

\begin{figure}[t]
    \includegraphics[width=0.48\textwidth, trim=20 0 0 0,  clip=yes]{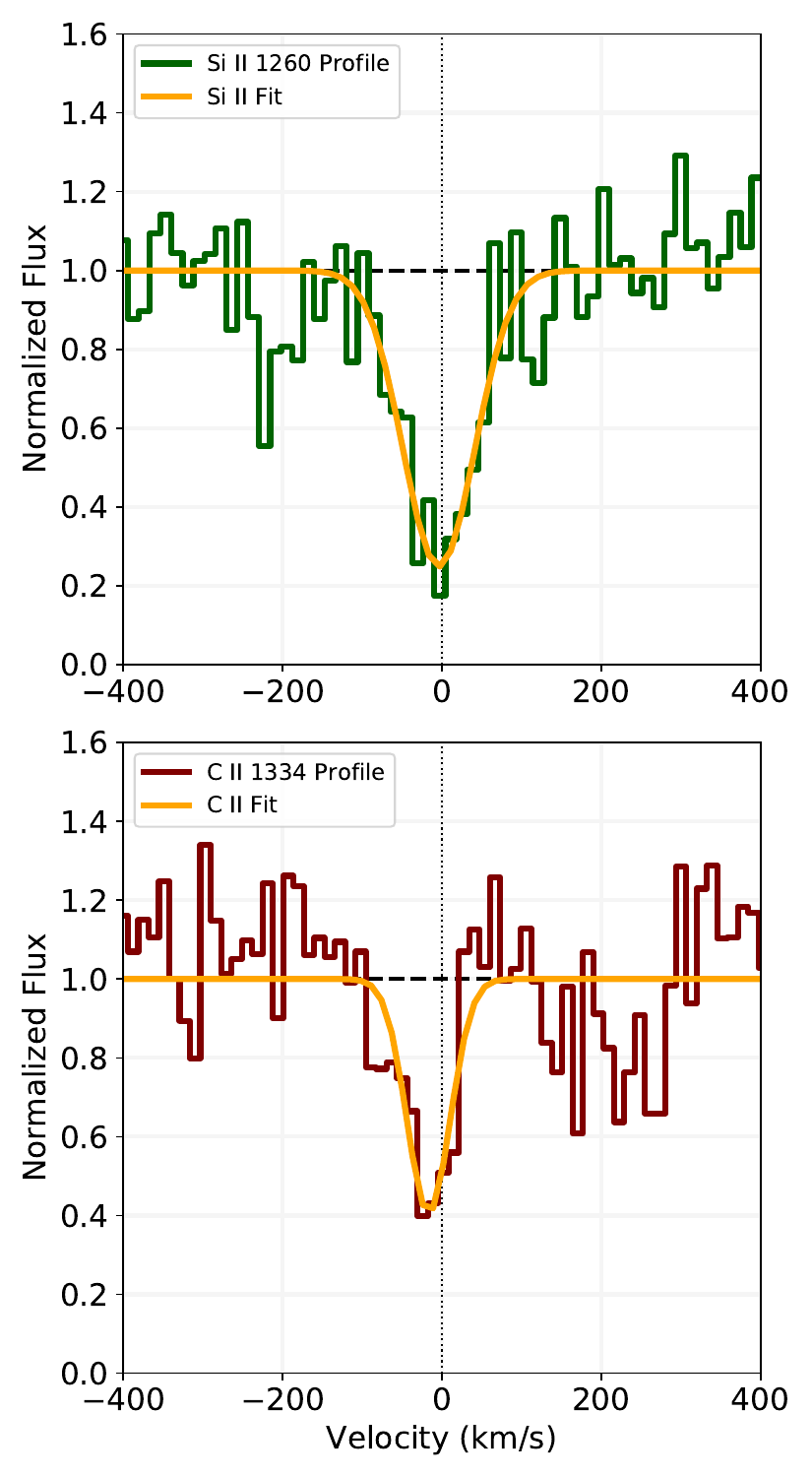}
    \caption{Low-ionization state absorption features present in Pox 186. \textit{Top:} \ion{Si}{2} 1260 \AA\ (green). The flux is normalized based on the continuum neighboring the absorption line. The fit to the absorption profile is provided in orange. \textit{Bottom:} The \ion{C}{2} 1334 \AA\ absorption profile and fit.}
    \label{fig:lis}
\end{figure} 

Further evidence of neutral gas is found in the presence of low-ionization state absorption features in the UV spectrum: in the top left panel of Figure \ref{fig:compspec}, we denote the resonant absorption features of \ion{Si}{2} at 1190, 1193, and 1260 \AA\ with black dotted lines and non-resonant \ion{Si}{2}$^*$ 1265 \AA\ with a black dashed line. Other low-ionization state absorption features present in the composite COS spectrum include \ion{C}{2} 1334 and \ion{C}{2}$^*$ 1335, which are shown in the top right panel. These absorption lines are present in both the G130M spectrum and the newly acquired G160M spectrum, although the \ion{C}{2} 1334 profile in the new G160M spectrum is affected by the filter edge. The ionization potentials of Si and C are such that an amount of Si$^+$ and C$^+$ is present in the neutral gas surrounding Pox 186; deep \ion{Si}{2} and \ion{C}{2} absorption features, therefore, are indicative of high \ion{H}{1} column densities or neutral gas covering fractions \citep{heck2001,heck2011,leitet2013,jask2019}. We highlight the \ion{Si}{2} 1260 and \ion{C}{2} 1334 \AA\ profiles in Figure \ref{fig:lis}. The \ion{Si}{2}, \ion{C}{2}, and Ly$\alpha$ profiles in Pox 186 do not saturate, which could indicate a covering fraction less than unity through which this radiation can escape. However, a low column density of neutral gas combined with low gas-phase metallicity could produce the same observed LIS absorption trends. Assuming the depth of the absorption profiles is related to the covering fraction of neutral gas, the covering fractions measured from the \ion{Si}{2} and \ion{C}{2} features are 0.75$\pm$0.07 and 0.59$\pm$0.08, respectively.

Absorption by \ion{Si}{2}$^*$ and \ion{C}{2}$^*$ in the top two panels of Figure \ref{fig:compspec} reveals that the temperature of the primarily neutral medium is sufficient to excite Si$^+$ and C$^+$ ions to the first excited, fine-structure level such that absorption from this energy level can take place. In their analysis of high-ionization GP galaxies, \citet{jask2019} found that two GPs with $O_{32}$ $>$ 7 and significant Ly$\alpha$ absorption (J1335+0801 and J1448$-$0110) have strong \ion{Si}{2} and \ion{C}{2} absorption features. While the covering fractions we estimate for Pox 186 are smaller than those measured in J1335+0801 and J1448$-$0110 \citep[0.97$\pm$0.04 and 0.94$\pm$0.03, respectively,][]{mcKi2019}, the presence of low-ionization state and Ly$\alpha$ absorption features are indicative of a non-negligible column density of neutral gas along the line of sight.


The second row of Figure \ref{fig:compspec} reveals that the new G160M spectrum shows evidence of \ion{S}{4} $\lambda\lambda$1406,1417 emission (left) and faint [\ion{N}{4}]$\lambda$1483 emission (right). The ratio of [\ion{N}{4}]$\lambda$1483 to \ion{N}{4}]$\lambda$1486 is sensitive to the electron density, but the latter is not detected. While this pair is sensitive to very high densities, the observation that [\ion{N}{4}]$\lambda$1483 is more intense than \ion{N}{4}]$\lambda$1486 in Pox 186 is consistent with a low-density system. Furthermore, \ion{Si}{4} $\lambda$1403 is detected in absorption but the \ion{Si}{4} $\lambda$1394 profile is affected by noise in the continuum.

The third row plots two very-high ionization emission features observed in the G160M observations: double-peaked \ion{C}{4} $\lambda\lambda$1548,1551 (left panel), and faint \ion{He}{2} $\lambda$1640  (right panel). The presence of \ion{He}{2} and double-peaked \ion{C}{4} indicate that at least a portion of the gas can be described by a very-high ionization zone \citep{berg2021} containing ions such as He$^{2+}$ and C$^{3+}$. The \ion{C}{4} emission lines, like Ly$\alpha$, are resonant transitions; for Ly$\alpha$, a small velocity separation between the two peaks is indicative of low scattering and the escape of the resonantly scattered emission through low column densities of neutral gas \citep[e.g.,][]{verh2015}. While the emission from \ion{C}{4} is dependent on the abundance of C and the radiative transfer in the ionized gas, the small peak separation suggests that there is likely a non-zero escape of resonantly scattered radiation from the very-high ionization gas.

Finally, the bottom row plots the high signal-to-noise (S/N) \ion{O}{3}]$\lambda\lambda$1660,1666 doublet observed in the new G160M spectrum (left) and the intense \ciii$\lambda\lambda$1906,1909 emission measured in the archival G185M spectrum (right). The \ciii\ doublet can be used as a density diagnostic for the high-ionization zone and to determine the C/O relative abundance (see \S3.1). The \ion{Si}{3}]$\lambda$1892 emission line is also observed and could be used as a density diagnostic with \ion{Si}{3}]$\lambda$1883, but the latter falls in the G185M filter gap.

\subsection{Ly\texorpdfstring{$\alpha$}{a} and SPS Fitting}

We use stellar population synthesis (SPS) to fit the COS UV spectrum and obtain the stellar properties in Pox 186. We follow the fitting methodology of \citet{chis2019}: the model stellar continuum is assumed to be a combination of 50 \textsc{Starburst99} single-star models \citep{leit1999,leit2010,leit2014} sampling 10 ages (1-40 Myr) and 5 metallicities (0.05-2$\times$Z$_\odot$), with their intensities reddened by a stellar attenuation component. Each of the 50 models is assigned a weight in the fitting routine such that the best-fit spectrum is a weighted sum of all the models, and the attenuation component is calculated assuming a \citet{redd2016} UV attenuation law. We can determine the light-weighted age and metallicity of the inferred population of Pox 186 from the weight assigned to each model spectrum. We verify that the resolution of the \textsc{Starburst99} models, 0.4 \AA, is similar to that of the COS UV spectrum, which we measure to be 0.22 \AA\ from the Galactic absorption lines. Therefore, we do not smooth the COS spectrum before fitting.

\begin{figure*}[ht]
   \centering
   \includegraphics[width=0.95\textwidth, trim=0 0 0 0,  clip=yes]{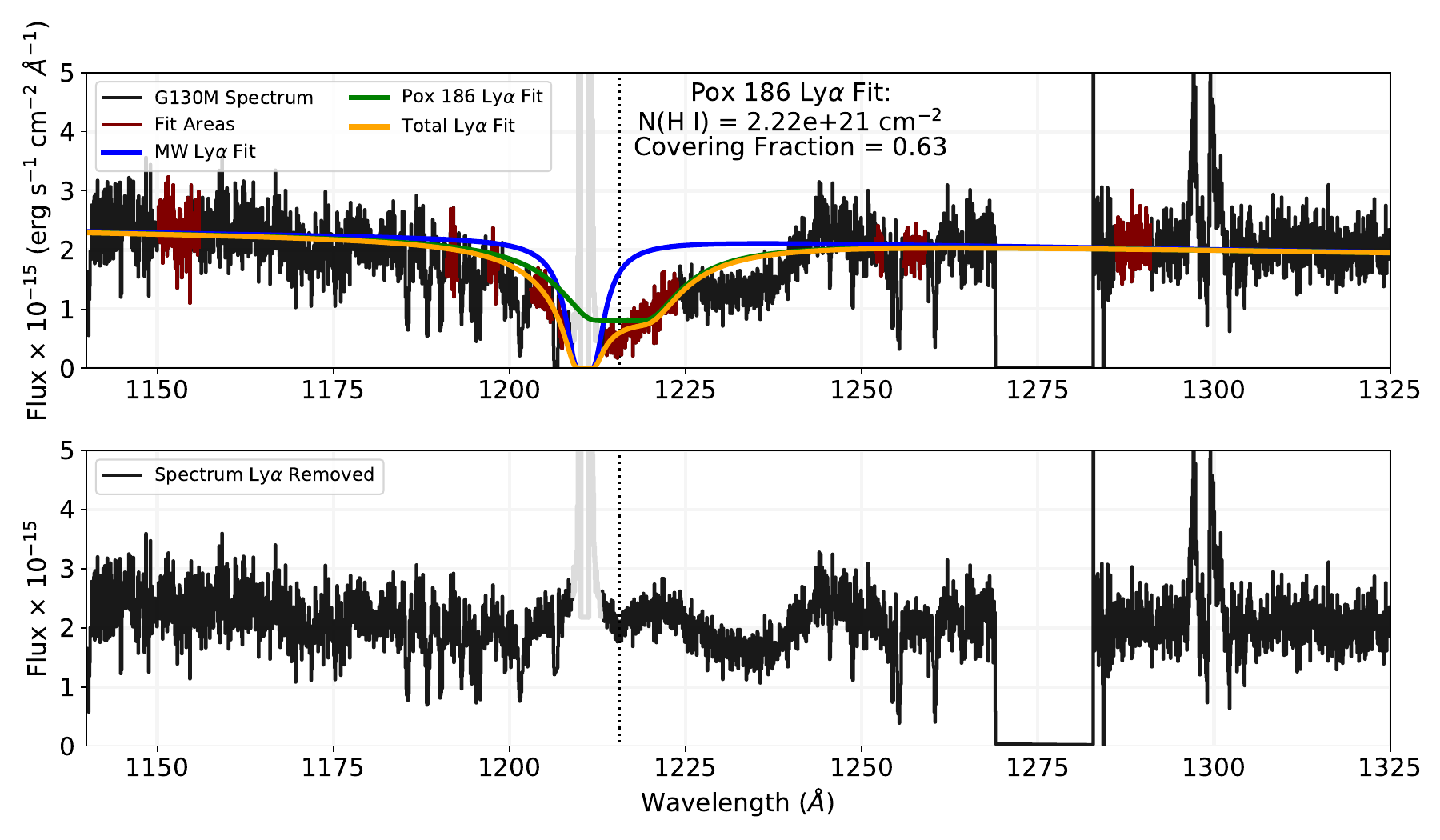}
   \caption{Portions of the G130M spectrum containing the Galactic, geocoronal, and Pox 186 Ly$\alpha$ profiles. \textit{Top:} Spectrum (black), the Galactic Ly$\alpha$ absorption profile determined from Galactic extinction (blue), and the Pox 186 Ly$\alpha$ profile (green) determined from the areas of the spectrum free from other spectral features (red). The combination of Galactic and Pox 186 Ly$\alpha$ absorption is shown in orange. Geocoronal Ly$\alpha$ is grayed out. \textit{Bottom:} The spectrum with the composite Ly$\alpha$ profile removed.}
   \label{fig:lya_prof}
\end{figure*} 

Certain spectral features are sensitive to the metallicity and age of the stellar population, requiring that these features be uncontaminated for SPS modeling. One such feature is the \ion{N}{5} $\lambda$1240 P Cygni profile, the presence of which indicates young (Age $<$ 10 Myr) populations \citep[see Figure 1 in][]{chis2019}. The \ion{N}{5} $\lambda$1240 P Cygni profile is blended with both Galactic and Pox 186 Ly$\alpha$ absorption, requiring that we deblend all three components before fitting the continuum. First, we use the \textsc{Python dustmaps} program \citep{gree2018} to determine the foreground reddening from the \citet{schl1998} dust maps. The magnitude of Galactic extinction, E(B$-$V)=0.046, is used with the \citet{card1989} reddening law and a R$_V$ of 3.1 to correct the UV spectrum for foreground extinction. Next, we model the Galactic Ly$\alpha$ absorption assuming a covering fraction of unity and a Doppler parameter of 30 km s$^{-1}$. We determine Galactic N(\ion{H}{1}), the column density of neutral H, using the relation from \citet{bohl1978} and the foreground E(B$-$V).

For the Pox 186 Ly$\alpha$ profile, we fit for both N(\ion{H}{1}) and the covering fraction assuming a Doppler parameter of 30 km s$^{-1}$ \citep[consistent with previous analyses,][]{proc2015,hu2023}. In doing this fit, we take precaution to avoid the wings of geocoronal Ly$\alpha$, any narrow absorption lines, and the \ion{N}{5} P Cygni profile. The top panel of Figure \ref{fig:lya_prof} plots the G130M spectrum (black), the areas we use for Ly$\alpha$ fitting (red), and the resulting Ly$\alpha$ profiles: Galactic in blue, Pox 186 in green, and the net Ly$\alpha$ profile in orange. From the Ly$\alpha$ fit, we estimate a \ion{H}{1} column density of N(\ion{H}{1}) $\gtrsim$ 2.2$\times$10$^{21}$ cm$^{-2}$ in Pox 186 and a neutral gas covering fraction of $\sim$0.63, in good agreement with the covering fraction estimated from \ion{C}{2} 1334 absorption profile (see Figure \ref{fig:lis}). We repeat this analysis using the foreground extinction from \citet{schl2011}, E(B$-$V)=0.039, and find that N(\ion{H}{1}) and neutral gas covering fraction are robust to the lower extinction.

\begin{figure*}[ht]
   \centering
   \includegraphics[width=0.85\textwidth, trim=40 0 40 0,  clip=yes]{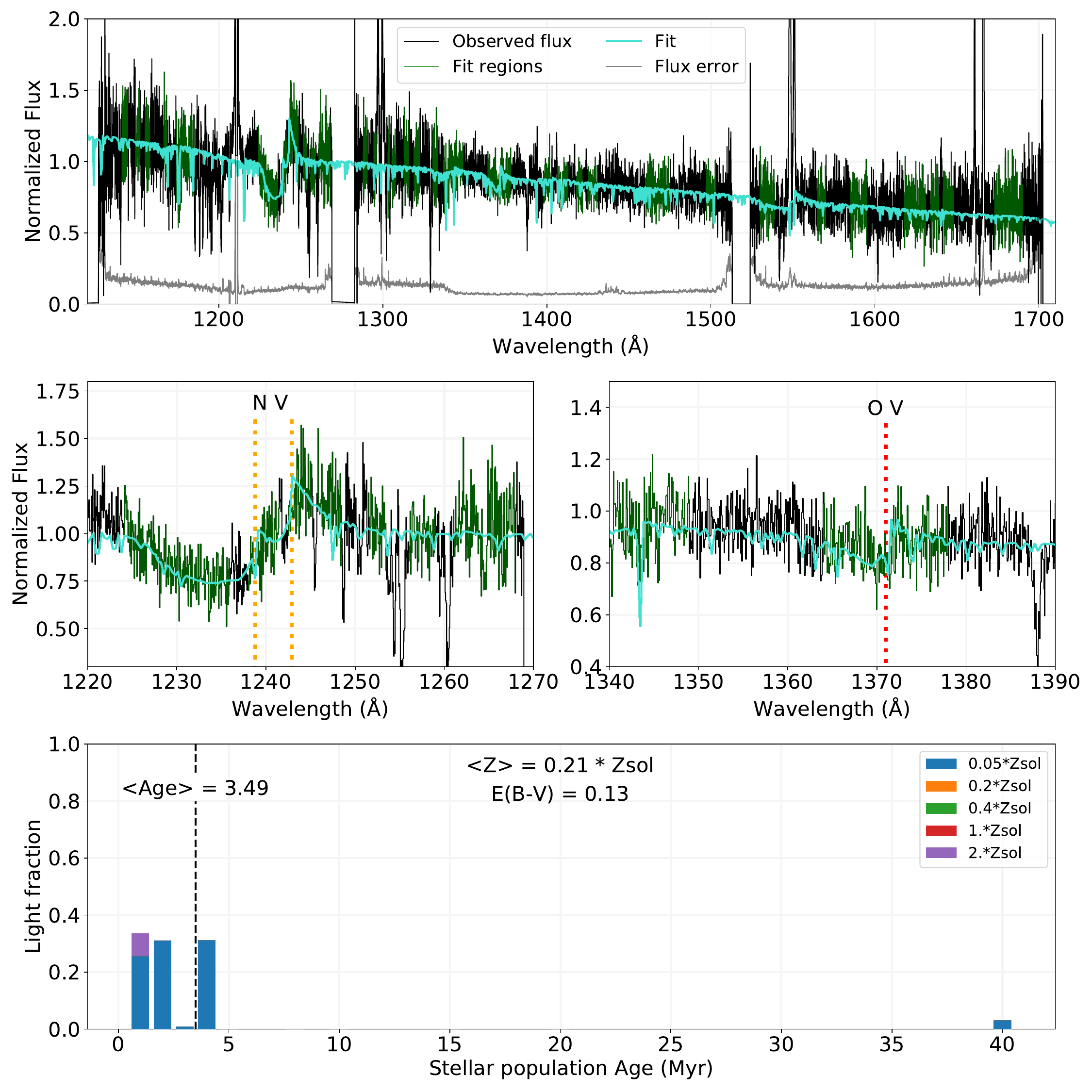}
   \caption{SPS results for Pox 186. \textit{Top Row:} Full spectrum after correction for Galactic and Pox 186 Ly$\alpha$ absorption. The spectrum is normalized between 1256 and 1268 \AA, the regions used for fitting are highlighted in green, and the best-fit model continuum is provided in cyan. \textit{Middle Row:} Two stellar wind features, \ion{N}{5} (left) and \ion{O}{5} (right) are highlighted. These are sensitive to the age of the stellar population. \textit{Bottom Row:} The fraction of each model in the best-fit continuum, where model ages are plotted on the x-axis and model metallicities are provided in different colors. The best-fit age, metallicity, and E(B$-$V) are given in this panel.}
   \label{fig:sps_res}
\end{figure*}

The final G130M spectrum after Galactic and Pox 186 Ly$\alpha$ removal is provided in the bottom panel of Figure \ref{fig:lya_prof}. We combine this spectrum with the new G160M spectrum at 1337 \AA\ to perform the SPS analysis. The G130M+G160M spectrum provides the wavelength coverage to fit key spectral components and constrain the reddening, and we take care to mask any MW or ISM absorption features before fitting. We also normalize the G130M+G160M and model spectra by the median flux between 1256 and 1268 \AA\ before fitting. The top panel of Figure \ref{fig:sps_res} plots the normalized UV spectrum without Ly$\alpha$ absorption (black), the areas used for fitting (green), and the best-fit SPS continuum (cyan). As shown in the middle panels, the models can fit the observed \ion{N}{5} profile and there is evidence of a faint \ion{O}{5} P Cygni profile. The \ion{O}{5} wind feature has been observed in the stacked spectra of extreme GPs \citep{jask2019} and is indicative of young, very massive stars with large ionizing photon production \citep[e.g.,][]{smit2016}.

The fits using the full range of models and the provided portions of the UV spectrum result in a light-weighted age, metallicity, and stellar attenuation of $\langle$Age$\rangle = 3.49$ Myr, $\langle$Z$_*\rangle = 0.21 \times$Z$_\odot$, and E(B$-$V) = 0.129, respectively, as exhibited in the bottom panel of Figure \ref{fig:sps_res}. The composite spectrum is comprised of $\gtrsim$90\% young (Age $\leq$ 4 Myr), 0.05$\times$Z$_\odot$ models. The inclusion of the 2$\times$Z$_\odot$ models is likely required to match the depth of the \ion{N}{5} P Cygni profile, which increases with metallicity at fixed age. The depth of the \ion{N}{5} P Cygni profile is sensitive to the removal of Ly$\alpha$ absorption; as such, deviation from the true Ly$\alpha$ profile in Pox 186 could produce larger \ion{N}{5} absorption/a slightly larger $\langle$Z$_*\rangle$. We obtain uncertainties on the light-weighted properties using a bootstrap approach: the input UV spectrum (i.e., the green portions of the spectrum in the top panel of Figure \ref{fig:sps_res}) is resampled a number of times equal to the number of elements in the original spectrum, then the new spectrum is fit with the stellar models to generate the light-weighted properties. We repeat this 100 times and determine the average and standard deviation of each distribution. The average properties measured from these distributions are Age$_{dist}$ = 3.35$\pm$1.06 Myr, Z$_{*,dist}$ = (0.21$\pm$0.03)$ \times$Z$_\odot$, and E(B$-$V)$_{dist}$ = 0.129$\pm$0.004, all of which are in good agreement with the SPS parameters determined from the fit to the original input spectrum. We assume that the standard deviations of the bootstrap resampled distributions are good approximations of the uncertainties on each property, such that we determine the light-weighted SPS properties of Pox 186 to be $\langle$Age$\rangle = 3.49\pm1.06$ Myr, $\langle$Z$_*\rangle = (0.21\pm0.03) \times$Z$_\odot$, and E(B$-$V) = 0.129$\pm$0.004.

The small contribution of the 40 Myr models ($\sim$5\%) does produce both a larger light-weighted age and uncertainty on the age estimates, and the high equivalent width (EW) of H$\beta$ \citep[$>$ 330 \AA,][]{guse2004} would suggest a young ionizing population in Pox 186. We repeat this analysis using only models with Age $\leq$ 10 Myr, and we find that the light-weighted metallicity and reddening remain unchanged while the light-weighted age is $\langle$Age$_{<10}\rangle = 2.33\pm0.11$ Myr. Both $\langle$Age$\rangle$ and $\langle$Age$_{<10}\rangle$ are consistent with the age estimate of 4$^{+1}_{-1.6}$ Myr and 3 Myr that \citet{corb2002} determine from a fit to Pox 186's optical continuum and EW of H$\beta$, respectively.

We also fit the UV continuum using the Binary Population And Spectral Synthesis \citep[BPASS,][]{eldr2017,stan2018} v2.2.1 model continua which account for the presence of binary populations. We use models with the same ages and metallicities as the \textsc{Starburst99} models, and we fit the same portions of the UV continuum highlighted in Figure \ref{fig:sps_res}. The resulting best-fit metallicity and reddening are in good agreement with the previous results: $\langle$Z$_{BPASS}\rangle = (0.25\pm0.04) \times$Z$_\odot$, and E(B$-$V)$_{BPASS}$ = 0.115$\pm$0.007. However, the best-fit age ($\langle$Age$_{BPASS}\rangle = 11.7\pm1.4$ Myr) is biased high because the continuum including Ly$\alpha$ is not provided when performing the fitting: this area of the continuum is relatively uncertain owing to the mixed geocoronal, Galactic, and Pox 186 Ly$\alpha$ profiles. The BPASS model spectra show significant Ly$\alpha$ absorption for older populations, and we do not observe this in the G130M+G160M spectrum (see Figure \ref{fig:lya_prof}). Limiting the models to those with Age $<$ 10 Myr produces a light-weighted age of 2.9$\pm$0.2 Myr with no change to the average metallicity or E(B$-$V), consistent with the \textsc{Starburst99} modeling results. In summary, our SPS results indicate that a significant fraction of the stars responsible for producing the highly-ionized ISM in Pox 186 have formed recently and, like the gas in this galaxy, are metal-poor.

\subsection{Emission Line Fits}

We fit the UV emission lines assuming a Gaussian profile with a local linear continuum. We assume that emission lines of the same ion in close proximity have the same Gaussian FWHM and offset from vacuum wavelength. The notable exception to this general fitting schematic is the profile for \ion{C}{4}. To obtain an estimate of the \ion{C}{4} $\lambda\lambda$1548,1551 fluxes, we fit the double-peaked emission lines in two ways. The first method is identical to that described in \citet{berg2019L}: we fit \ion{C}{4} $\lambda$1548, the stronger of the two lines, with two narrow Gaussians to describe the emission towards the blue and red side of the line center. We then use these Gaussian fits and the emissivity ratio of \ion{C}{4} $\lambda$1548 and \ion{C}{4} $\lambda$1551 \citep[j$_{1548}$/j$_{1551}$ = 1.97 from the atomic transition probabilities of][]{wies1996} to infer the fits to \ion{C}{4} $\lambda$1551. For the second method, we fit both \ion{C}{4} $\lambda$1548 and \ion{C}{4} $\lambda$1551 with the two Gaussian components. We compare both methods and find that the second method produces slightly larger fluxes for \ion{C}{4} $\lambda$1551. This is similar to what \citet{berg2019L} find for the EELG J141851 (see their Figure 3). We use the double Gaussian fits for both \civ$\lambda$1548 and \civ$\lambda$1551 in this work since both lines are detected at high S/N and we can accurately fit \civ$\lambda$1551. We model the P Cygni \ion{C}{4} profiles in a separate analysis (Scarlata et al. in preparation).

\heii$\lambda$1640 is detected at a S/N of 2.8 in the new G160M observations that we present, and the emission is very faint (see Figure \ref{fig:compspec}). In fitting \ion{He}{2} $\lambda$1640, we note that the FWHM of the single Gaussian fit is broader than the other nebular UV emission lines (including the neighboring \oiiisf\ lines), which \citet{corb2002} also observe in the STIS spectrum of Pox 186. Broad \ion{He}{2} emission is usually associated with a stellar origin from Wolf-Rayet (WR) stars, but stellar \ion{He}{2} emission is typically observed in higher metallicity sources \citep{senc2017}. \citet{nana2019} examine a sample of $z \sim$2-4 galaxies that have nebular \ion{He}{2} emission and find that the average FWHM of \ion{He}{2} $\lambda$1640 is also broader than the other UV lines. \citet{guse2004} detect \ion{He}{2} $\lambda$4686 in the MMT and 3.6m ESO telescope optical spectra of Pox 186, but they do not mention any evidence of significant stellar \ion{He}{2}. Additionally, GMOS IFU data obtained by \citet{egge2021} confirm the presence of narrow, nebular \ion{He}{2} $\lambda$ 4686 with no stellar contribution. As such, we constrain the width of \ion{He}{2} $\lambda$1640 using the neighboring \ion{O}{3}] strong lines to avoid overestimating the flux of this faint line. We assess the validity of this fit in \S3.2.

\begin{deluxetable}{lcccc}
\tablewidth{\textwidth}
\tabletypesize{\footnotesize}
\tablecaption{Pox 186 COS Emission Line Intensities}
\tablehead{
  \colhead{Ion}  & 
  \colhead{Obs. $\lambda$}  & 
  \colhead{Filter}  & 
  \colhead{$\frac{I(\lambda)}{I(\mbox{O III]})}$}  &
  \colhead{EW (\AA)}}
\startdata
\ion{S}{4}   &  1406.03   &  G160M  &  0.048 $\pm$ 0.009  &  0.19 $\pm$ 0.03 \\
\ion{S}{4}   &  1416.90   &  G160M  &  0.025 $\pm$ 0.009  &  0.10 $\pm$ 0.03 \\
\ion{N}{4}   &  1483.32   &  G160M  &  0.041 $\pm$ 0.008  &  0.17 $\pm$ 0.04 \\
\ion{C}{4}   &  1548.18   &  G160M  &  0.905 $\pm$ 0.027  &  4.15 $\pm$ 0.10 \\
\ion{C}{4}   &  1550.77   &  G160M  &  0.508 $\pm$ 0.023  &  2.34 $\pm$ 0.10 \\
\ion{He}{2}  &  1640.41   &  G160M  &  0.047 $\pm$ 0.017  &  0.24 $\pm$ 0.09 \\
\ion{O}{3}]   &  1660.80   &  G160M  &  0.339 $\pm$ 0.015  &  1.76 $\pm$ 0.07 \\
\ion{O}{3}]   &  1666.14   &  G160M  &  1.000 $\pm$ 0.022  &  5.23 $\pm$ 0.08 \\
\ion{Si}{3}]  &  1892.02   &  G185M  &  0.376 $\pm$ 0.060  &  3.42 $\pm$ 0.55 \\
\ion{C}{3}]   &  1906.62   &  G185M  &  2.470 $\pm$ 0.079  &  20.16 $\pm$ 0.57 \\
\ion{C}{3}]   &  1908.67   &  G185M  &  1.610 $\pm$ 0.077  &  13.15 $\pm$ 0.59 \\
\hline
\hline
E(B - V)     &      &   0.129$\pm$0.004   &     &     \\
F(O III])    &      &   5.36$\pm$0.08   &     &     \\
\enddata
\label{t:poxInt}
\tablecomments{Line intensities measured from the COS spectrum of Pox 186. The columns are: 1. The ion that produces the emission line; 2. Observed line center (\AA); 3. COS filter the emission line is observed in; 4. Line intensity relative to \ion{O}{3}] $\lambda$1666; 5. The EW (in \AA) of the emission line. The line intensities are determined by correcting the line fluxes with the E(B$-$V) obtained from the SPS fitting. This E(B$-$V) is reported in the line second from the bottom, and the flux of \ion{O}{3}] $\lambda$1666 (in units of 10$^{-15}$ ergs s$^{-1}$ cm$^{-2}$) before reddening correction is reported in the last line.}
\end{deluxetable}

To correct the spectrum for dust extinction, we use the \citet{redd2016} UV extinction law with E(B$-$V)=0.129$\pm$0.004 (as determined from the SPS model fits, see \S2.2); the SPS E(B$-$V) is in agreement with the attenuation inferred from the optical Balmer line decrements, although we note that prior works have found a large range for the nebular E(B$-$V) in Pox 186 \citep[from $\sim$0 to 0.3, see discusison in][]{egge2021}. The emission line intensities are reported relative to the intensity of \oiiisf$\lambda$1666 in Table \ref{t:poxInt}. Table \ref{t:poxInt} also reports the EW of the emission lines, which reveals the large EWs of the \oiiisf, \ciii, and \civ\ lines. Particularly, we measure a combined EW of \ciii$\lambda\lambda$1906,1909 $>$ 33 \AA, significantly larger than the EW(\ciii) observed in GPs \citep[$<$ 10 \AA,][]{ravi2020}, z$>$6 galaxies \citep[up to 22 \AA,][]{star2017}, and comparable to the largest EW(\ciii) measured in SF dwarf galaxies in the local universe \citep[$>$30 \AA,][]{ming2022}.

\section{Physical Conditions}

We now highlight the physical conditions in the extreme ISM of Pox 186 that we can measure from the composite COS UV spectrum. When necessary, we use $O_{32}=$ 18.3$\pm$0.1 and 12+log(O/H) $=$ 7.74$\pm$0.01 dex from \citet{guse2004} and $T_e$\oiii\ $=$ 16,920$\pm$450 K from \citet{egge2021}.

\subsection{\texorpdfstring{$n_e$(C III]), C$^{2+}$/O$^{2+}$, and C/O Abundance}{ne(C III]), C2+/O2+, and C/O Abundance}}

The \ciii$\lambda\lambda$1907,1909 doublet is sensitive to the electron density, $n_e$, in the high-ionization gas containing C$^{2+}$, O$^{2+}$, and other species. The traditional density diagnostic in the optical is the \sii$\lambda\lambda$6717,6731 doublet, but this ion is present in the low-ionization zone of the ISM which only describes a small fraction of the gas in Pox 186 as indicated by the very large $O_{32}$. \citet{guse2004} measure a $n_e$(\sii) of 70$^{+110}_{-70}$ cm$^{-3}$ from the MMT spectrum which resolves both lines in the doublet; the two \sii\ lines are blended in the 3.6m ESO telescope optical spectrum and produce a larger density of $n_e$(\sii)$=$350$\pm$60 cm$^{-3}$.

We calculate the electron density in Pox 186 using the \textsc{Python PyNeb} package \citep{luri2012,luri2015L}. The \textsc{getTemDen} function is used with $T_e$\oiii\ and the C$^{2+}$ atomic data from \citet{glas1983}, \citet{nuss1978}, and \citet{wies1996} and collision strengths from \citet{berr1985}. We determine that the electron density in Pox 186 as measured from the \ciii\ lines is in the low-density limit. The \ciii$\lambda$1907/\ciii$\lambda$1909 ratio is 1.53$\pm$0.08, consistent with the line emissivity ratio for a gas at the provided $T_e$\oiii\ and $n_e \sim$ 1 cm$^{-3}$: j$_{1907}$/j$_{1909}$=1.53. The problem with constraining $n_e$(\ciii) is that the critical densities of \ciii$\lambda$1907 and \ciii$\lambda$1909 are such that the emissivity ratio is most sensitive to electron densities between 10$^4$ and 10$^5$ cm$^{-3}$. Using the minimum observed line ratio, we constrain the $n_e$(\ciii) upper limit to be $n_e$(\ciii)$<$2300 cm$^{-3}$ such that the denisty in Pox 186 is $n_e$(\ciii)$=$1$^{+2300}_{-1}$ cm$^{-3}$. Since $n_e$(\sii) and $n_e$(\ciii) both suggest that the electron densities in Pox 186 are in the low-density limit, we use $n_e$(\ciii)$=$1 cm$^{-3}$ for all abundance calculations. We have repeated the following abundance calculations at $n_e$ $=$ 70, 350, and 2300 cm$^{-3}$ but find that all densities produce the same ionic abundances.

Using the intensity of the strong \ciii, \civ, and \oiiisf\ UV lines and the emissivity of the transitions, we calculate the relative C$^{2+}$/O$^{2+}$ and C$^{3+}$/O$^{2+}$ ionic abundances. The uncertainty on each ionic abundance is determined through a propagation of uncertainties in the emission line fluxes and on $T_e$\oiii. We measure the ionic abundances as C$^{2+}$/O$^{2+}$ $=$ 0.220$\pm$0.008 and C$^{3+}$/O$^{2+}$ $=$ 0.039$\pm$0.001. Despite the high-ionization environment in Pox 186, it cannot be assumed that all C is in the doubly- and triply-ionized states or that all O is in the doubly-ionized state. There are no easily-observed C$^+$ emission lines in the UV or optical, so we must employ an Ionization Correction Factor (ICF) to account for the unobservable C in the ISM. We use the ICF(C) of \citet{berg2019}, which was developed using \textsc{CLOUDY} photoionization models \citep{ferl2013} with varying input 12+log(O/H), log(C/O), and stellar ages and metallicities. This ICF is parameterized as a third-order polynomial function in log($O_{32}$) to determine log(U), the ionization-parameter. The inferred log(U) is then used in a fourth-order polynomial to determine ICF(C) such that C/O $=$ ICF(C)$\times$(C$^{2+}$/O$^{2+}$). The polynomial fits are dependent on the metallicity of the ISM; we utilize the 10\% solar metallicity ICFs to match the gas-phase metallicity of Pox 186 (11\% Z$_{\odot}$).

Using the \citet{berg2019} ICF, we measure log(U) $=$ $-$1.93 and log(C/O)$_{ICF}$ $=$ $-$0.62$\pm$0.02. This is slightly different from the log(C/O) abundance measured using only the UV C and O lines: log(C/O)$_{Sum}$ $\approx$ log(C$^{2+}$/O$^{2+}$ + C$^{3+}$/O$^{2+}$) $=$ $-$0.59$\pm$0.01. However, if we use the optical O$^{+}$/O$^{2+}$ ratio from \citet{guse2004} to account for the missing O$^+$ we find: log(C/O)$_{Otot}$ $=$ log((C$^{2+}$ + C$^{3+}$)/(O$^{2+}\times$(1+O$^{+}$/O$^{2+}$))) $=$ $-$0.61$\pm$0.01 when using the MMT and 3.6m abundance ratio. Provided that log(C/O)$_{Otot}$ is consistent with log(C/O)$_{ICF}$, we adopt the latter when discussing the C/O abundance in Pox 186. This comparison would imply that most of the C in the gas-phase of Pox 186 is in the C$^{2+}$ or C$^{3+}$ states, but that there is significant O$^{+}$ to be accounted for when determining the C/O abundance.

The C/O abundance in Pox 186 is in good agreement with C/O measured in local low-metallicity dwarf starburst galaxies, which span a range of $-$1.05 $<$ log(C/O) $<$ $-$0.38 \citep{berg2019}. It is difficult to compare to C/O abundances in more distant, similarly high ionization galaxies owing to a lack of UV \oiiisf\ and \ciii\ emission line detections. \citet{ravi2020} report on the UV observations of ten GP galaxies; they detect \ciii\ emission in seven galaxies and use the blended emission lines to determine C/O using UV \oiiisf\ and optical \oiii\ lines. While the C/O abundance in Pox 186 does agree with the range of abundances determined in the GPs ($-$1.329 to $-$0.580 when using the upper limits on \oiiisf), the latter are uncertain due to low S/N detections of the \ciii\ and \oiiisf\ lines and lack of sufficient spectral resolution to deblend the individual emission line components. The scatter in C/O abundance, particularly at low metallicities, is related to the star formation history of a galaxy, including starburst duration, retention of oxygen in the ISM, and the number of starbursts that have taken place. \citet{berg2019} show that the C/O abundances measured in local dwarf starburst galaxies can be reproduced by chemical evolution models dependent on these initial conditions. We do not try to model these in Pox 186; regardless, despite the extreme ionization conditions and UV emission line EWs, the gas-phase C/O abundance in Pox 186 suggests a star formation history that is consistent with other starburst dwarf galaxies in the local universe.

\subsection{UV He II Emission}

\begin{figure*}[ht]
   \centering
   \includegraphics[width=0.9\textwidth, trim=40 0 40 0,  clip=yes]{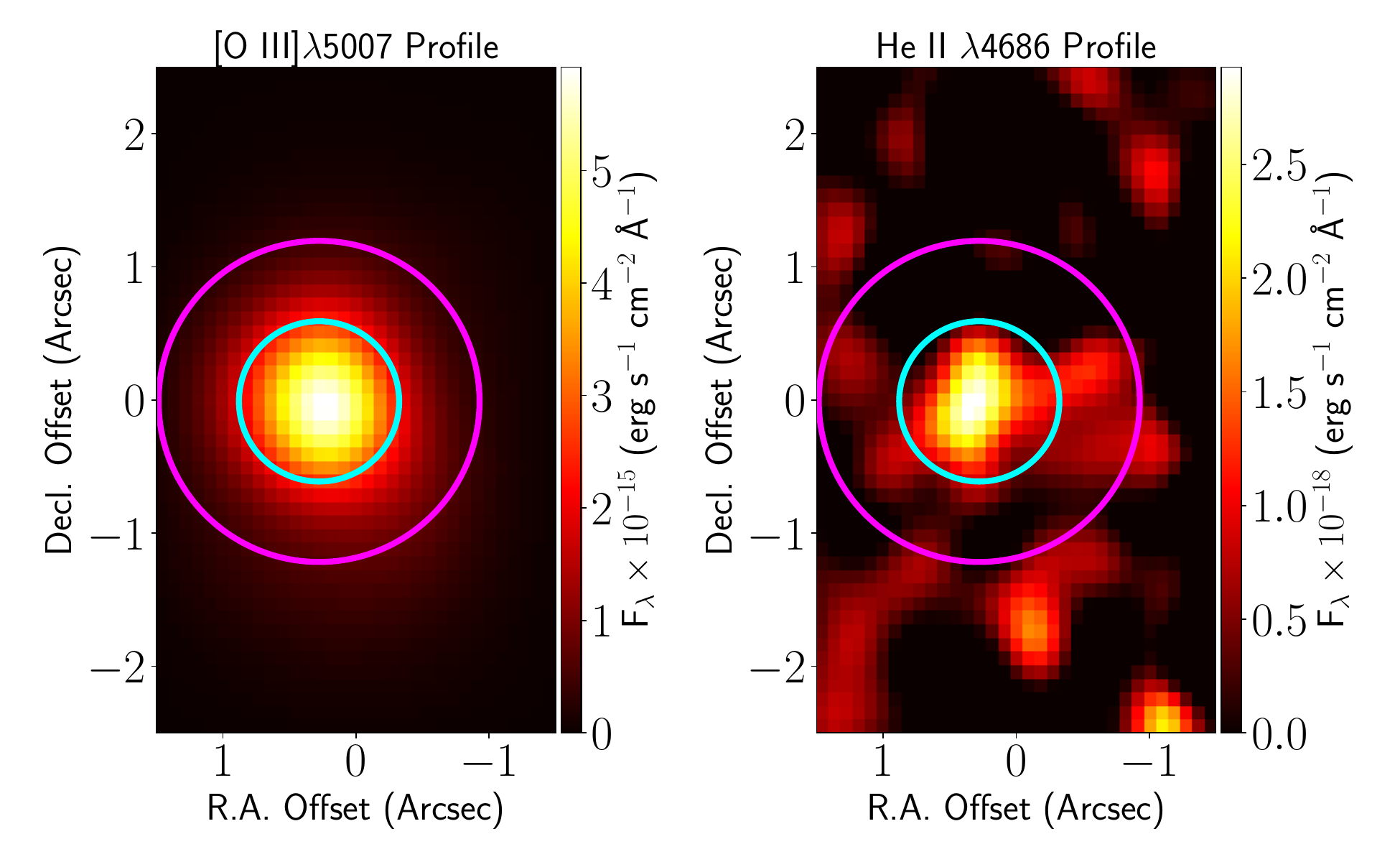}
   \caption{The Pox 186 \oiii$\lambda$5007 (left) and \heii$\lambda$4686 (right) spatial distributions obtained from the GMOS IFU data. The magenta extraction profile matches the 2\farcs5 diameter of the COS beam, while the cyan profile mirrors the narrower, un-vignetted 1\farcs25 diameter beam. The \oiii$\lambda$5007 emission fills the narrow beam and extends into the full 2\farcs5 aperture while the \heii\ emission is compact.}
   \label{fig:gmos}
\end{figure*} 

The \heii$\lambda$1640 emission line indicates the production of very high energy photons that are able to doubly-ionize He. The UV \heii\ emission in Pox 186 is significantly fainter than the emission observed in other EELGs; therefore, it is imperative to assess the validity of this fit given this line's utility in UV emission line diagnostics \citep[e.g.,][]{felt2016}. This is possible with the intensity and emissivity of the optical \heii\ and \oiii\ lines.

To verify our fit to \heii$\lambda$1640, we employ the optical GMOS IFU spectrum of Pox 186 obtained by \citet{egge2021}. For the GMOS IFU observations, we extract the optical spectra using a circular aperture with a 2\farcs5 diameter to match the nominal COS aperture and a 1\farcs25 diameter to simulate a narrow, un-vignetted COS aperture. Figure \ref{fig:gmos} plots the distribution of the continuum-subtracted \oiii$\lambda$5007 and \heii$\lambda$4686 emission. While the \oiii\ emission fills the 1\farcs25 aperture and extends into the 2\farcs5 aperture, the \heii\ emission is compact. This is confirmed when examining the spatial profiles of both \oiii$\lambda$5007 and \heii$\lambda$4686: the FWHM of the \oiii\ and \heii\ emission from the GMOS spaxels are 1\farcs06 and 0\farcs83, respectively. Provided that \heii\ emission traces very-high ionization gas, this indicates that the ionization structure in Pox 186 follows the traditional \hii\ region structure where the very-high energy photons photoionize the gas in close proximity to the ionizing source. We note that the only significant structure in the right panel is the peak \heii\ emission at the center of the 1\farcs25 aperture, and the faint emission off center is dominated by noise in the continuum.

We extract the total spectrum within each circular aperture of Figure \ref{fig:gmos} and measure the flux of \heii$\lambda$4686 with a single Gaussian. For \oiii$\lambda$5007, we fit a Gaussian plus Lorentzian profile to account for the broad wings produced by the outflow in Pox 186 \citep{egge2021}. We then measure the \heii/\oiii\ ratio using only the narrow component of \oiii$\lambda$5007, and multiply this by the emissivity ratio $\frac{j_{5007}\times j_{1640}}{j_{4686}\times j_{1666}}$ to infer the UV \heii/\oiiisf\ ratio. In fitting the emission lines we note that, for both apertures, the FWHM of \heii$\lambda$4686 is less than the FWHM of \oiii$\lambda$5007; this would favor narrow, nebular \heii\ emission as opposed to the broad emission that WR stars produce. Provided that these lines are in close proximity, we do not apply a reddening correction to the optical spectra before measuring the line ratios. We measure F(4686)/F(5007) $=$ (3.3$\pm$0.7)$\times$10$^{-5}$ and (3.9$\pm$0.5)$\times$10$^{-5}$ for the 2\farcs5 and 1\farcs25 diameter apertures, respectively. These produce expected UV intensity ratios of I(1640)/I(1666) $=$ 0.049$\pm$0.010 and 0.059$\pm$0.007, respectively; both are consistent with the intensity ratio reported in Table \ref{t:poxInt} of I(\heii)/I(\oiiisf) $=$ 0.047$\pm$0.017, indicating that the \heii$\lambda$1640 measured in the COS G160M spectrum is similar to what is expected from the optical \heii\ and \oiii\ lines. If we allow for a broad Gaussian fit to \heii$\lambda$1640, the I(\heii)/I(\oiiisf) ratio we measure (0.074$\pm$0.022) is still in agreement with the expected ratios from the optical. Provided that we do not observe broad \heii$\lambda$4686, we proceed with the narrow fit to \heii$\lambda$1640 and note that the interpretations of the emission line ratios in \S4 are robust to the lower UV emission line intensity.

We apply the above procedure to the \citet{guse2004} optical emission line intensities from the MMT and 3.6m ESO spectra to explore the ionization structure of Pox 186 as probed by IFUs/larger beams and single-object longslits. We find that the expected UV line ratios are 0.12$\pm$0.05 and 0.27$\pm$0.02 for the MMT and 3.6m spectrum, respectively. These ratios are larger than the I(\heii)/I(\oiiisf) measured in the COS G160M spectrum and inferred from the GMOS IFU data. One potential issue is that these apertures are not directly comparable to the COS aperture: the \citet{guse2004} MMT longslit spectrum is extracted in a 2\farcs$\times$6\farcs\ area while the 3.6m spectrum is extracted in a 1\farcs$\times$3\farcs6 area but oriented at a position angle of $-$68$^{\circ}$. For the MMT spectrum, the large extraction area will include much of the significant \oiii$\lambda$5007 emission but can introduce more noise to the continuum which can affect the fits of particularly faint lines like \heii$\lambda$4686. As for the 3.6m spectrum, the extraction area is between that of the full and un-vignetted COS aperture but is focused on the peak emission in Pox 186. A slit with width 1\farcs\ is smaller than the spatial FWHM of the \oiii$\lambda$5007 emission in Pox 186 but is larger than the spatial FWHM of \heii$\lambda$4686. Such an aperture may omit a significant portion of the \oiii$\lambda$5007 flux while including the majority of the \heii\ emission, thereby producing an enhanced \heii/\oiii\ ratio and implying a much larger production of very-high energy photons capable of He$^+$ ionization.

Local objects are most susceptible to significant differences introduced by aperture size. Recently, \citet{arel2022classy} investigated SF galaxies with multiple observations from longslits, IFUs, and circular apertures (including COS) and find that physical properties such as $T_e$, $n_e$, and metallicity are relatively insensitive to the choice of aperture. However, physical parameters from the very-high ionization zone containing \heii\ emission are not considered in the analysis. Most of the physical conditions examined are predominantly determined from ratios of emission lines originating in the same ionization zone (such as $T_e$ and $n_e$) or that are relative to the ubiquitous H$^+$ emission in the \hii\ region (like O$^{2+}$/H$^+$ and E(B$-$V)). As long as these line ratios remain relatively constant within the given ionization zone, then there should be little variation in the physical conditions as measured from different apertures. This is not the case when comparing emission lines from different ionization zones, such as \heii$\lambda$4686 and \oiii$\lambda$5007. If the ionization structure of other SF dwarf galaxies is similar to that of Pox 186, then the \heii\ emission will be included in all apertures centered on the peak emission. Emission from the high-ionization zone containing O$^{2+}$ could vary depending on the aperture selected (like the 3.6m spectrum discussed above), which will most notably change the I(\heii)/I(\oiii) ratio since the \heii\ emission does not extend significantly into this ionization zone (see Fig \ref{fig:gmos}). Given these aperture differences, we do not believe that the optical \heii/\oiii\ line ratios from \citet{guse2004} are directly comparable to the COS G160M line ratios and maintain the earlier conclusion that the COS G160M \heii/\oiiisf\ ratio is appropriate for emission line diagnostics in Pox 186.

The low \heii/\oiii\ ratios are not purely a product of intense \oiii\ emission in Pox 186, but are related to faint \heii\ emission: the optical \heii/H$\beta$ ratio is (2.0$\pm$0.4)$\times$10$^{-3}$ and (2.5$\pm$0.3)$\times$10$^{-3}$ for the 2\farcs5 and 1\farcs25 aperture GMOS IFU extractions, respectively. The \heii/H$\beta$ ratio is measured between (8-26)$\times$10$^{-3}$ for GP galaxies \citep{jask2013,fern2022} and LyC leakers \citep{guse2020}, and the lowest ratios in the SDSS SF galaxies are $\sim$3.5$\times$10$^{-3}$ \citep{shir2012}. While the source of intense \heii\ emission is still up for debate, the faint emission measured in Pox 186 is consistent with the trends of photoionization models for regions ionized predominantly by stellar populations and with little contribution from other more energetic sources like high-mass x-ray binaries \citep[HMXBs, see models from][]{senc2020} or AGN (ruled out from UV line ratios, see \S4). The ability to detect faint optical and UV \heii\ in Pox 186 is likely due to its proximity, which is supported by the larger H$\beta$ fluxes measured in the GMOS data relative to the extreme GPs or other EELGs. Only nine of the original 80 SF GPs from \citet{card2009} have significant \heii\ detections \citep{hawl2012}, but a higher fraction of SF galaxies could have faint \heii\ that is undetected due to the combination of lower fluxes/larger noise in the continuum. While many studies have focused on the intense \heii\ emission in SF systems \citep[e.g.,][]{kehr2015,berg2021}, it is equally important to understand the limits of \heii\ emission and how the production of very high energy photons evolves in SF galaxies.

\section{Emission Line Diagnostics}

\subsection{UV Line Ratios and Photoionization Models}

We now focus on the UV emission line ratios observed in Pox 186 to understand the ionization conditions of this extreme starburst galaxy. Many have proposed various line diagnostics in the UV, but we only analyze the ratios with the detected lines of \ciii, \civ, \oiiisf, and \heii. For comparison, we draw upon a sample of galaxies from the literature with significant UV emission line measurements. This sample includes the compact dwarf galaxies used to study carbon and oxygen abundances from \citet{berg2016}, nearby SF regions selected for optical \ion{He}{2} emission in the SDSS from \citet{senc2017}, metal-poor dwarf galaxies from \citet{berg2019}, and the galaxies comprising the COS Legacy Archival Spectroscopic SurveY \citep[CLASSY,][]{berg2022} with the UV emission line intensities for each galaxy reported by \citet{ming2022}. When there is overlap between these samples, we default to the CLASSY line intensities.

\begin{figure*}[ht]
   \centering
   \includegraphics[width=0.9\textwidth, trim=40 0 40 0,  clip=yes]{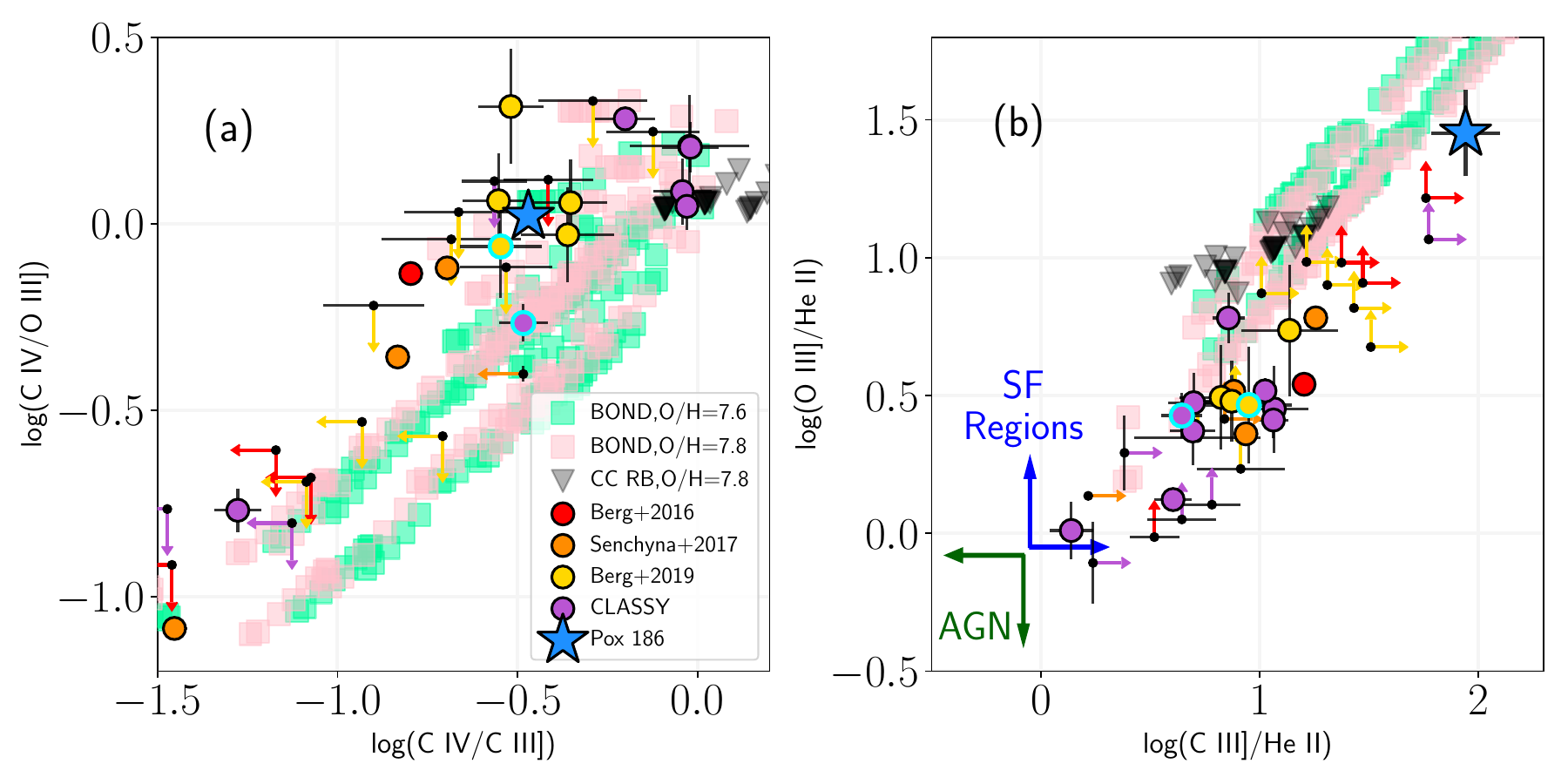}
   \caption{The UV emission line ratios measured in Pox 186 (blue star), literature SF galaxies (circles, various colors with limits indicated by black dots with arrows of the same color), and photoionization models from BOND (lightly colored squares) and nonequilibrium models (gray triangles). The photoionization models are selected to have metallicity and log(U) similar to that of Pox 186. \textit{Panel (a):} log(\civ/\oiiisf) vs.\ log(\civ/\ciii); \textit{Panel (b):} log(\oiiisf/\heii) vs.\ log(\ciii/\heii). Panel (b) can distinguish between stellar ionizing sources (ratios in the area noted by blue arrows) and AGN (noted by green arrows). Two SF galaxies discussed in \S4.2.2 are highlighted with cyan borders.}
   \label{fig:uv_lines}
\end{figure*} 

Figure \ref{fig:uv_lines} plots the line ratios for the literature sources (various colored circles) and for Pox 186 (blue star). Lower and upper limits on the literature line ratios are represented by black points with colored arrows (colored to match the literature sample the limits are taken from). In this figure, \oiiisf\ = I(\oiiisf$\lambda$1660) + I(\oiiisf$\lambda$1666), \ciii\ = I(\ciii$\lambda$1906) + I(\ciii$\lambda$1909), \civ\ = I(\civ$\lambda$1548) + I(\civ$\lambda$1551), and \heii\ = I(\heii$\lambda$1640). We utilize photoionization models to assess the position of Pox 186 and the SF galaxies in this figure. We start with the Bayesian Oxygen and Nitrogen abundance Determination \citep[BOND,][]{vale2016} models available from the Mexican Million Models Database \citep[3MdB,][]{mori2015}. These models span a large range in physical conditions, namely 2.8 dex in 12+log(O/H), 4 dex in input log(U), different geometries, and radiation- and density-bounded systems. We focus on a subset of models selected for their ability to reproduce \hii\ region and blue-compact galaxy (BCG) emission line spectra. These criteria are discussed in \citet{amay2021} and specifically focus on a range in 12+log(O/H), log(N/O), log(U), and the area of the \oiii/H$\beta$ vs.\ \nii/H$\alpha$ diagram \citep[proposed by][or BPT]{bald1981} where the \hii\ region and BCGs are observed. However, we do not apply any limits on the starburst ages and we consider all radiation- and density-bounded models because there is evidence that Pox 186 may be a density-bounded system \citep{guse2004,egge2021}. We further restrict the models by only considering those with metallicity and log(U) similar to those measured in Pox 186; this limits the models to those with 7.6$\leq$12+log(O/H)$\leq$7.8 and $-$2.15$\leq$log(U)$<$$-$1.65. The UV line ratios obtained from the resulting BOND models are plotted as light green (12+log(O/H)=7.6) and pink (12+log(O/H)=7.8) squares in Figure \ref{fig:uv_lines}.

\citet{egge2021} found that there is high-velocity gas in Pox 186 indicative of outflows. The temperature structure in high-velocity winds can deviate from adiabatic assumptions, which can result in different cooling functions in the wind \citep[e.g.,][]{sili2004}. This ``catastrophic cooling" has been used to explain the observed optical emission line ratios of extreme SF galaxies, including the GPs \citep{jask2019}. Recently, \citet{dane2022} studied the effects of radiative cooling in nonequilibrium ionization states and produced photoionization model UV emission line intensities for different systems characterized by strong galactic outflows. These models are parameterized as a function of metallicity, mass loading rate, wind velocity, and ambient density. We include the UV emission line ratios from the lowest metallicity (Z=0.125$\times$Z$_\odot$, or 12+log(O/H)$\sim$7.8), radiation-bounded models from \citet{dane2022} as gray triangles in Figure \ref{fig:uv_lines} and labeled as ``CC RB".

Panel (a) plots log(\civ/\oiiisf) against log(\civ/\ciii), ratios that are sensitive to the C/O abundance and the ionization structure in the gas, respectively. We find that the ratios measured in Pox 186 are in good agreement with the SF galaxies from the literature, particularly the \citet{berg2019} metal-poor dwarf galaxies (yellow circles). The metallicity and C/O relative abundance measured in Pox 186 is consistent with the C/O abundances measured in these galaxies, which explains the similar log(\civ/\oiiisf) ratios. The consistent log(\civ/\ciii) ratios also indicate that the ionization structures in Pox 186 and the metal-poor dwarf galaxies are similar, but this is slightly dependent on how the \civ\ lines are fit in Pox 186. As noted in \S2.3, we use the double Gaussian fit to the \civ\ lines despite the presence of a P Cygni profile. In the absence of self-absorption, the \civ$\lambda\lambda$1548,1551 intensity in Pox 186 would increase to larger log(\civ/\oiiisf) and log(\civ/\ciii), and place Pox 186 in a similar area of the diagram as the CLASSY galaxies (purple circles).

The \civ/\oiiisf\ and \civ/\ciii\ line ratios measured in Pox 186 and the literature SF galaxies are consistent with the BOND low-metallicity photoionization models. Furthermore, the literature upper limits all suggest line ratios that occupy the same area of parameter space covered by the models. We note that the range of metallicity and log(U) chosen is likely not applicable for all the literature galaxies and that extending the limits on either variable can produce agreement between the literature and BOND model ratios. The CC RB model ratios involving the \civ\ lines are offset to higher values relative to both the BOND model ratios and to most of the SF galaxies in the literature sample. This is consistent with the findings in \citet{dane2022}: generally, the predominant spectroscopic changes in the nonequilibrium ionization state models are in the very high-ionization species of \ion{O}{6} and \ion{C}{4} and the emission lines from O$^{2+}$ and C$^{2+}$ show little change from standard photoionization models. While the \civ\ line ratios measured in Pox 186 are different from the \citet{dane2022} model ratios, the environment in Pox 186 might be distinct from those assumed in the models (e.g., lower metallicity or lower stellar mass).

Panel (b) plots log(\oiiisf/\heii) against log(\ciii/\heii), which has been shown to separate SF systems and AGN: photoionization models with AGN spectral energy distributions (SEDs) have log(\ciii/\heii) and log(\oiiisf/\heii) $\lesssim$ 0 while the same ratios in SF systems are typically $\gtrsim$ 0 \citep{felt2016}. This can be understood in terms of the hardness of an AGN ionizing spectrum, which will produce more emission from highly-ionized species (like \heii) relative to SF systems. As can be seen, the literature galaxies and Pox 186 clearly fall in the SF area of the diagnostic diagram and agree with the BOND photoionization model line ratios. The photoionization models span a broad range of log(\oiiisf/\heii) and log(\ciii/\heii), extending to values $>$ 2.5 in each ratio.

The \heii\ emission line ratios measured in Pox 186 are obviously distinct from those of the literature SF galaxies. This difference is partially a selection effect, as some of the literature galaxies were selected for UV observations based on their optical \heii\ emission. Other samples, such as CLASSY, more generally target compact SF galaxies spanning a broad range of parameter space, yet none of the robustly measured log(\oiiisf/\heii) or log(\ciii/\heii) line ratios are observed at $>$ 1.2. The intrinsically low \heii\ flux in Pox 186 produces the only robustly measured log(\oiiisf/\heii) and log(\ciii/\heii) ratios in this area of the diagnostic diagram, requiring both faint \heii\ and intense \oiiisf\ and \ciii\ emission. While there is a dearth of measured ratios $\gtrsim$ 1.2, the literature lower limits indicate that faint \heii\ emitters would occupy the same area of the line ratio diagram.

\subsection{Exploring the UV Trends in Pox 186}

What, then, could produce the faint \heii\ and large \oiiisf/\heii\ or \ciii/\heii\ ratios observed in Pox 186? We explore two potential explanations for a low output of He$^+$ ionizing photons in this otherwise very-high ionization galaxy: age of the stellar population and total stellar mass.

\begin{figure*}[ht]
   \centering
   \includegraphics[width=0.9\textwidth, trim=40 0 40 0,  clip=yes]{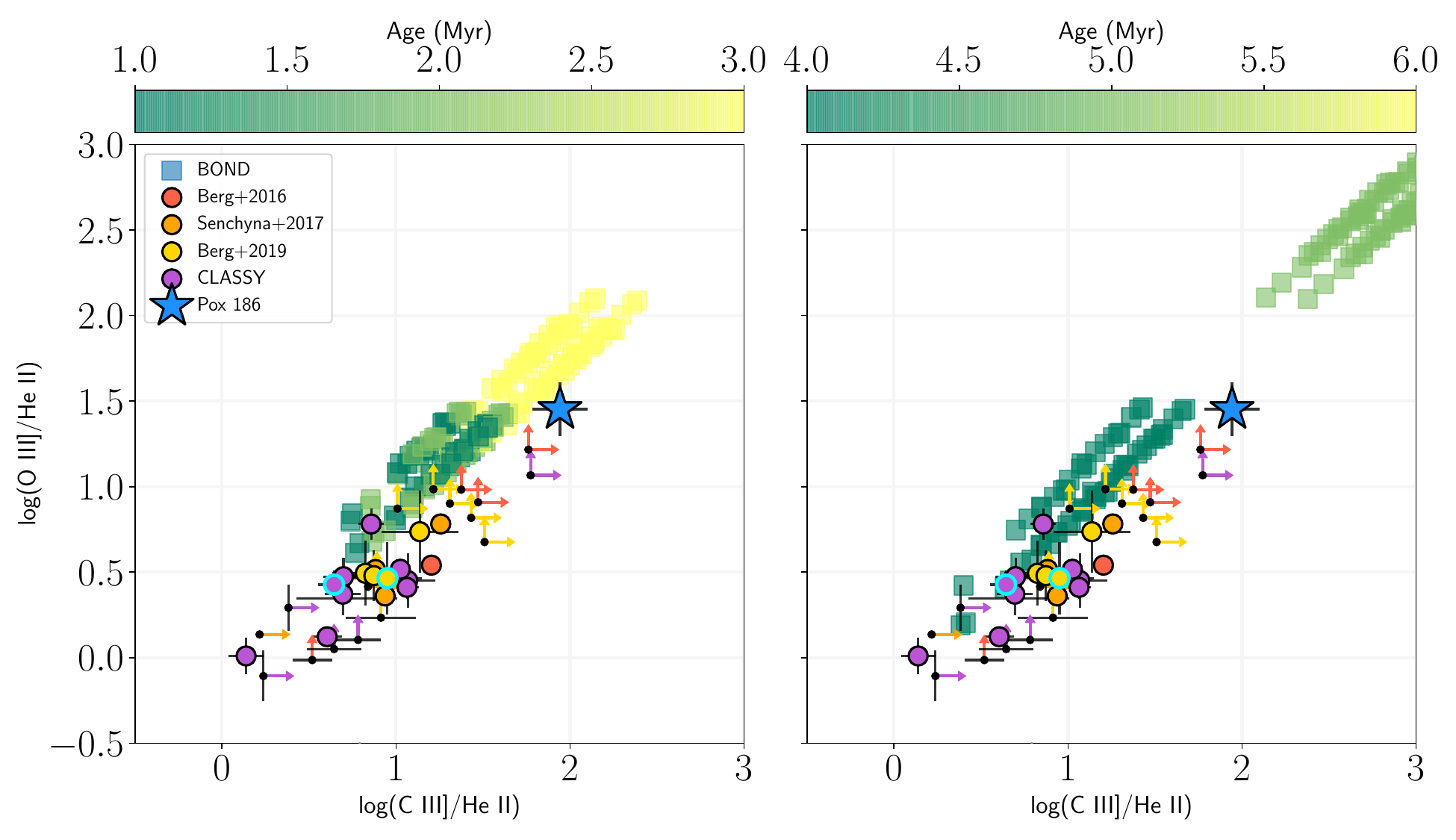}
   \caption{log(\oiiisf/\heii) vs.\ log(\ciii/\heii) with the same data as Panel (b) of Figure \ref{fig:uv_lines}, except the BOND photoionization model line ratios are color-coded by SED age. The model line ratios cover Age ranges from 1-3 Myr (\textit{Left Panel}) and 4-6 Myr (\textit{Right Panel}).}
   \label{fig:uv_lines_age}
\end{figure*} 

\subsubsection{Age of the Population}

The production of very-high energy photons from a low-metallicity ionizing stellar population decreases with age as the most massive stars exit the main sequence. At higher metallicities, older stellar populations enter a WR phase that can increase the hardness of the ionizing spectrum and produce intense emission from high-ionization species. While SPS fitting indicates that the population in Pox 186 is both metal-poor and young, a larger contribution of older stars could explain the faint \heii\ emission measured in both the optical and UV. The BOND models are constructed using the PopStar SEDs \citep{moll2009} sampled in 1 Myr steps from 1-6 Myr. We now explore the evolution of the UV emission line ratios/the hardness of the ionizing continuum in the photoionization models at different ages.

In Figure \ref{fig:uv_lines_age}, we replot the log(\oiiisf/\heii) vs.\ log(\ciii/\heii) ratios of Pox 186, the literature SF galaxies, and the BOND models, color-coding by SED age. We maintain the same 12+log(O/H) and log(U) ranges from Figure \ref{fig:uv_lines}. The left panel focuses on the SEDs with Age $\leq$ 3 Myr while the right panel examines the SEDs with Age $\geq$ 4 Myr. We have increased the plot range to study the evolution of \heii\ emission as a function of age. We find that log(\oiiisf/\heii) and log(\ciii/\heii) are both sensitive to the age of the ionizing population. Within this metallicity range, the ionizing spectra of 1-2 Myr stars produce log(\oiiisf/\heii) and log(\ciii/\heii) between $\sim$0.5 and 1.25, consistent with the ratios and limits of the literature SF galaxies. As age increases, the \heii\ emission decreases: the 3 Myr models predict line ratios that agree with those measured in Pox 186 and extend to \heii\ emission that is $<$1\% of the strong UV \oiiisf\ and \ciii\ lines. The hardness of the ionizing spectrum decreases further at 5 and 6 Myr, where the latter models produce line ratios that extend beyond the upper plot limits in the right panel of Figure \ref{fig:uv_lines_age}.

The only deviation from the trend of increasing population age and log(\oiiisf/\heii) or log(\ciii/\heii) is found in the 4 Myr models. At metallicities 7.6$\leq$12+log(O/H)$\leq$7.8, the PopStar SEDs enter a brief WR phase at $\sim$4 Myr that results in a hard ionizing spectrum. This increases the \heii\ emission and yields UV line ratios that are roughly equivalent or less than the 1-2 Myr model ratios. The enhanced \heii\ emission from the onset of the WR phase is the reason why there is a large gap in parameter space in the right panel of Figure \ref{fig:uv_lines_age}. The increased \heii\ emission can be sufficient to match the low log(\oiiisf/\heii) observed in some of the literature SF galaxies.


To summarize, nebular UV \heii\ emission is sensitive to the age of the ionizing population: intense ($\gtrsim$5\% \oiiisf\ or \ciii\ intensity) \heii\ is driven by young (1-2 Myr) stellar populations with hard ionizing spectra, while faint ($\leq$5\% \oiiisf\ or \ciii\ intensity) \heii\ emission indicates an older (Age $\geq$ 3 Myr) population with a softer ionizing spectrum. The onset of the WR phase can enhance \heii\ relative to the other high-ionization UV emission lines, but this can be assessed by the width of \heii$\lambda$1640 and other spectral features such as the broad optical blue and red WR features from \ion{C}{3}, \ion{C}{4}, and \heii\ \citep{lope2010}. However, the utility of \heii\ emission as an age indicator is predicated on the assumption that the \heii\ emission is nebular in origin. Other ionizing sources that can power \heii\ emission in SF galaxies include shocks, HMXBs, and the soft x-ray emission from superbubbles, the last of which can produce He$^+$ ionizing photons for clusters up to an age of 20 Myr \citep[see discussion in][]{oski2022}.

\begin{figure}[t]
    \includegraphics[width=0.48\textwidth, trim=20 0 0 0,  clip=yes]{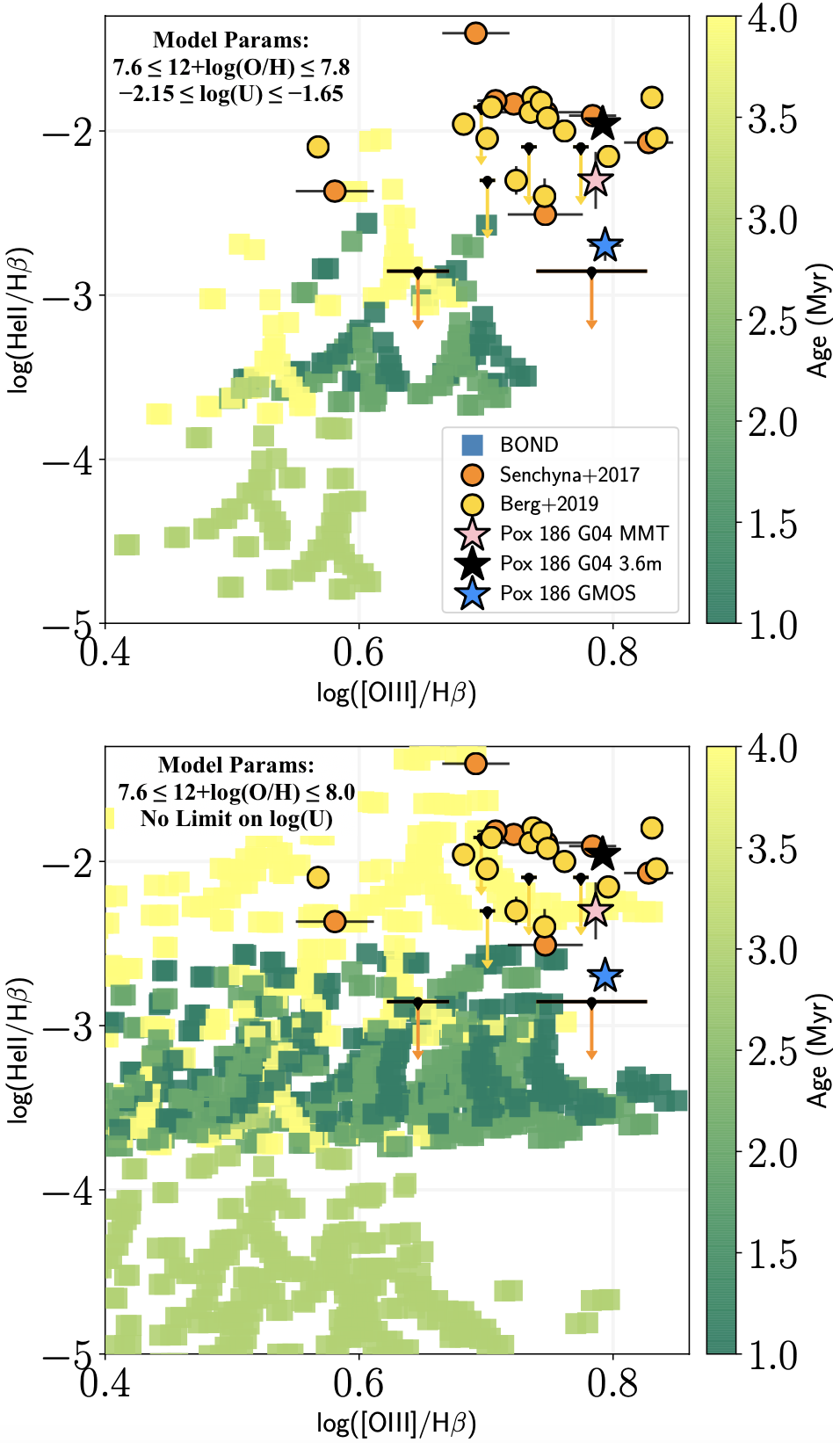}
   \caption{The optical line ratios log(\heii$\lambda$4686/H$\beta$) vs.\ log(\oiii$\lambda$5007/H$\beta$). For Pox 186, the black and pink star represent the \citet{guse2004} MMT and 3.6m ESO spectra, respectively, while the blue star indicates the ratios measured from the GMOS IFU spectrum. The optical ratios from literature SF galaxies appear as colored circles and limits. \textit{Top Panel:} The same BOND photoionization models used in Figure \ref{fig:uv_lines}, excluding the 5 and 6 Myr models. \textit{Bottom Panel:} The metallicity range of the photoionization models is increased and no limits are placed on log(U).}
   \label{fig:optical_lines_age}
\end{figure}

These other sources of ionization are often invoked to explain the discrepancy between measured and modeled \heii$\lambda$4686 emission in the optical. To see this, we plot the \heii$\lambda$4686/H$\beta$ and \oiii$\lambda$5007/H$\beta$ ratios for the literature SF galaxies and Pox 186 in Figure \ref{fig:optical_lines_age}. We note that \citet{berg2016} do not report the intensity of \heii$\lambda$4686 and the optical spectra of the CLASSY galaxies are currently not reported in a homogeneous manner. The different observations of Pox 186 are noted as the various colored stars. We also plot the BOND line ratios for models with SED Age 1-4 Myr, excluding the 5 and 6 Myr models because they have significantly lower \heii/H$\beta$ ratios and cannot reproduce the UV emission line ratios from the literature. From the top panel, it is clear that the optical ratios from the literature and in Pox 186, even those measured from the GMOS IFU spectrum, occupy an area of the diagram that the photoionization models do not predict. The models in this panel are selected using the same metallicity and log(U) ranges as Figure \ref{fig:uv_lines}, which do not reflect the physical conditions in some of the literature SF galaxies but should be characteristic of those in Pox 186.

In an attempt to match the observed line ratios in the sample of SF galaxies, we extend the BOND model parameter space to 7.6$\leq$12+log(O/H)$\leq$8.0 and place no limits on log(U). The results are plotted in the bottom panel of Figure \ref{fig:optical_lines_age}. To achieve log(\heii/H$\beta$) $>$ $-$2.5 requires 4 Myr models/the WR phase, and none of the 1-2 Myr model SEDs can reach the intense optical \heii\ emission observed in the literature SF galaxies. While the 1-2 Myr SEDs predict \heii/H$\beta$ ratios consistent with the GMOS IFU observations of Pox 186, these models simultaneously fail to reproduce \oiii/H$\beta$.

Taken together, the UV and optical emission line diagrams may indicate that the photoionization models assume a simplified picture of the stellar population in these SF galaxies. One assumption of the PopStar models is that the ionizing population is composed of stars at a fixed age/metallicity and only considers single-star models. For Pox 186, the SPS approach in \S2.2 uses a linear combination of stellar spectra and finds that a mixture of stars with ages $\leq$ 4 Myr best match the stellar features in the UV continuum. This mixture of stellar ages could explain the need for both 1-2 Myr and 4 Myr stars to simultaneously match the UV and optical emission line ratios, respectively, of the literature dwarf galaxies. Furthermore, binary stars can produce harder ionizing spectra than single-star models at larger ages as they extend the lifetime of the WR and other later phases of evolution \citep{stan2016}. Binary populations may better explain the lowest log(\oiiisf/\heii) ratios in the literature sample; for example, the CLASSY observations of J0337$-$0502 find that log(\oiiisf/\heii) and log(\ciii/\heii) are $\sim$0, which require metal-rich 4 Myr PopStar models that have entered the WR phase. However, the spectrum of this object shows intense, narrow \heii\ that would argue against a significant WR component to the \heii\ emission and for a harder ionizing spectrum from a different source. Taken together, a significant population of stars with age $\geq$ 3 Myr could be responsible for faint \heii\ emission in SF galaxies, but a lack of systems with faint \heii\ limits the comparison to photoionization model line ratios.

\subsubsection{Low Stellar Mass}

Another potential explanation for the faint \heii\ emission in Pox 186 is a sparsely sampled Initial Mass Function (IMF) due to a low total stellar mass. To see this, we focus on the line ratios of dwarf galaxies observed by \citet{berg2019} with O/H, C/O, $z$, log(\civ/\oiiisf), and log(\civ/\ciii) similar to those measured in Pox 186 and with significantly detected \heii$\lambda$1640 emission. This limits the comparison to two dwarf galaxies: J095430 and J141851. The latter of these galaxies is included in the CLASSY sample and is identified therein as J1418+2102; for this galaxy, we use the emission line ratios and physical conditions reported by \citet{ming2022} for the comparison to Pox 186. The UV morphologies of both J095430 and J141851 \citep[see Figure 1 of][]{berg2019} are similar to Pox 186 in that they are resolved but still fit within the COS beam. The line ratios measured in these two dwarf galaxies are highlighted with bold cyan borders in Figures \ref{fig:uv_lines} and \ref{fig:uv_lines_age}.

Table \ref{t:poxComps} compares the galactic properties and physical conditions measured in each of these galaxies and in Pox 186. The properties of J095430 include z$_{J0954}$ = 0.005, 12+log(O/H)$_{J0954}$ = 7.70$\pm$0.02, and log(C/O)$_{J0954}$ = $-$0.60$\pm$0.11, all of which are in very good agreement with those of Pox 186 and explains the consistent line ratios in Panel (a) of Figure \ref{fig:uv_lines}. For J141851: z$_{J1418}$ = 0.009, 12+log(O/H)$_{J1418}$ = 7.75$\pm$0.02, and log(C/O)$_{J1418}$ = $-$0.89$\pm$0.07 using the CLASSY UV emission line fits and the \citet{berg2019} ICF. The significantly lower log(C/O) produces the difference in log(\civ/\oiiisf) in Figure \ref{fig:uv_lines} Panel (a), but the ionization structure as traced by log(\civ/\ciii) is comparable to that of Pox 186. We also provide $O_{32}$ for each galaxy, where $O_{32}$ for J095430 is approximated from the \oiii\ strong lines and \oii\ auroral lines; the $O_{32}$ for J141851 is reported in \citet{berg2021}. While $O_{32}$ in Pox 186 is larger than the two comparison galaxies, the agreement between the log(\civ/\ciii) ratios would suggest a similar ratio of very-high to high-ionization gas in these systems. Despite the agreement between the general properties and gas-phase conditions in all three dwarf galaxies, the \oiiisf/\heii\ and \ciii/\heii\ ratios measured in Pox 186 are significantly larger than those of J095430 or J141851 (Panel (b) of Figure \ref{fig:uv_lines}).

\begin{deluxetable*}{lcccccccc}
\tablewidth{\textwidth}
\tabletypesize{\footnotesize}
\tablecaption{Physical Conditions of Specific SF Galaxies}
\tablehead{
  \colhead{Name}  & 
  \colhead{$z$}  &
  \colhead{12+log(O/H)}  & 
  \colhead{log(C/O)}  & 
  \colhead{$O_{32}$}  & 
  \colhead{log(M$_*$/M$_\odot$)}  & 
  \colhead{log($L_{1500}$)}  &
  \colhead{log(L(H$\alpha$))}}
\startdata
Pox 186   &  0.0041  &  7.74$\pm$0.01  &   $-$0.62$\pm$0.02  &  18.3$\pm$0.1 & $\sim$5 &  25.4 &  39.501$\pm$0.007  \\
J095430  &  0.0050  & 7.70$\pm$0.02  &  $-$0.60$\pm$0.11  &  7.6$\pm$0.7 & 6.53$^{+0.09}_{-0.08}$   &  26.2 &  39.821$\pm$0.006  \\
J141851  &  0.0086  & 7.75$\pm$0.02  &  $-$0.89$\pm$0.07  & 4.7$\pm$0.1 &  6.22$^{+0.49}_{-0.35}$  &  26.2 &  40.086$\pm$0.006  \\
Leo P   &  0.0009  & 7.17$\pm$0.04  &  \nodata  & 3.1$\pm$0.1  &  5.75$^{+0.02}_{-0.18}$  &  23.1 &  36.466$\pm$0.015
\enddata
\label{t:poxComps}
\tablecomments{Comparison of the properties measured in Pox 186, two literature SF galaxies, and the extremely metal-poor galaxy Leo P. For each galaxy, the columns provide: 1. Galaxy name; 2. Redshift; 3. Gas-phase metallicity (dex); 4. log(C/O) relative abundance (dex) measured from the UV lines, where J141851's abundance is calculated from the \citet{ming2022} line ratios; 5. $O_{32}$ (inferred from \oii\ auroral lines for J095430);  6. Stellar mass; 7. The luminosity density at 1500 \AA\ (erg s$^{-1}$ Hz$^{-1}$); 8. H$\alpha$ luminosity (erg s$^{-1}$). The metallicity of Pox 186 is taken from the measurements of \citet{guse2004} while log(M$_*$) is reported by \citet{corb2002}.}
\end{deluxetable*}

The most notable difference between Pox 186, J095430, and J141851 is the stellar mass of each system: log(M$_{*,Pox 186}$/M$_\odot$) $\sim$ 5 \citep{corb2002}, log(M$_{*,J0954}$/M$_\odot$) = 6.53$^{+0.09}_{-0.08}$ as reported by the MPA-JHU database of stellar masses from SDSS data, and log(M$_{*,J1418}$/M$_\odot$) = 6.22$^{+0.49}_{-0.35}$ which \citet{berg2022} determine through the application of the Bayesian Analysis of Galaxy SED \citep[BEAGLE,][]{chev2016} code. Given the significantly lower stellar mass of Pox 186, the potential explanation for the weak \heii\ emission is that Pox 186 has not produced a sufficient number of high-mass, high-temperature stars to doubly ionize a significant fraction of He. This is supported by the the luminosity density at 1500 \AA, $L_{1500}$, determined from the COS UV spectrum of each galaxy. The intensity at 1500 \AA\ is free of emission line features and is set by the continuum of the massive stars within the galaxy. We determine the $L_{1500}$ as (F$_{1500}\times1500^2$/c)$\times4\pi$D$^2$ where F$_{1500}$ is the flux at 1500 \AA\ (erg s$^{-1}$ cm$^{-2}$ \AA$^{-1}$), c is the speed of light (\AA\ s$^{-1}$), and D is the distance to the galaxy (cm). As reported in Table \ref{t:poxComps}, $L_{1500}$ measured from the new HST spectrum of Pox 186 is almost an order of magnitude lower than $L_{1500}$ measured in the two comparison SF galaxies, consistent with Pox 186 having fewer high-mass stars to produce a strong UV continuum/the hard ionizing spectrum necessary for strong \heii\ emission.

To see the effects of a stochastically sampled IMF in a low mass system, we also include the extremely metal-poor galaxy Leo P in Table \ref{t:poxComps} for comparison. First discovered in \ion{H}{1} 21cm emission by \citet{giov2013}, this galaxy is isolated, nearby (D = 1.6$\pm$0.15 Mpc), has a low stellar mass of log(M$_{*,LeoP}$/M$_\odot$) = 5.75$^{+0.02}_{-0.18}$, and hosts one bright \hii\ region ionized by a single O star \citep{mcQu2015}. \citet{skil2013} obtained optical spectroscopy of this region and determined a gas-phase abundance of 12+log(O/H) = 7.17$\pm$0.04 and $O_{32}$ = 3.1$\pm$0.1, yet optical \heii\ was not detected. The lack of \heii$\lambda$4686 was confirmed by recent Keck Cosmic Web Imager IFU spectroscopy obtained by \citet{telf2023}, who complemented this optical IFU data with existing HST/COS UV spectroscopy of the O star and surrounding \hii\ region \citep{telf2021}. Supporting the findings in the optical, the UV spectrum of the single \hii\ region revealed \oiiisf\ in emission but no significant \civ\ or \heii\ emission lines. Without significant \heii\ or archival \ciii$\lambda\lambda$1906,1909 emission, the UV emission line ratios of Leo P cannot be included in Figure \ref{fig:uv_lines}. $L_{1500}$ measured from this spectrum is log($L_{1500}$) = 23.14 \citep{telf2023}, significantly lower than J095430 and J141851 and reflective of the single star that is producing the bulk of the emission at 1500 \AA.

The physical conditions in Pox 186 and Leo P are not exactly the same, with the most notable differences being the lower gas-phase metallicity and H$\alpha$ luminosity of Leo P, and the significant, albeit faint, \heii\ emission observed in Pox 186. Furthermore, Leo P is not a compact source: while the O star is a point source in the COS beam \citep{telf2021}, the \oiii\ emission fills the COS aperture and there are H$\alpha$ structures that extend well beyond the dominant \hii\ region \citep{evan2019}. Nevertheless, Pox 186 and Leo P reveal certain aspects about the ISM in very low-mass systems: 1. Although fewer in number, the stars in these objects are capable of ionizing the gas to a point where it is dominated by high-ionization species; 2. Their optical and UV spectra can be described by intense, large-EW emission lines comparable to those observed in more massive dwarf galaxies; 3. The production of very-high energy photons necessitates the presence of very massive stars, but a sufficient number of these stars may not be formed in the latest burst of star formation if the IMF is stochastically sampled \citep[see discussions in][]{lee2009,cerv2013}. Dwarf starburst galaxies with larger stellar masses can more reliably produce these stars given that they can fully sample the IMF, but low-mass systems may produce fewer of these stars (a potential explanation for Pox 186) or none at all (like Leo P).

A stochastically sampled IMF could reduce the number of He$^+$ ionizing photons and result in the faint \heii\ emission observed in Pox 186, similar to an aging population of stars. While a range of ages are examined in the BOND photoionization models and the age required for low log(\oiiisf/\heii) emission is consistent with SPS modeling, the PopStar models do not consider the effects of stochastic sampling when generating the SEDs. As \citet{stan2019} note, the total mass of the starburst has a significant impact on the production of very high energy photons, where the lack of a single very-massive star can greatly reduce the He$^+$ ionizing photon output. They estimate that a total starburst mass $>$ 10$^6$ M$_\odot$ is required to fully realize the IMF, while others have estimated that starburst masses below 10$^4$ M$_\odot$ may not reliably produce any stars with mass $>$ 100 M$_\odot$ \citep[e.g.,][]{elme2000,daSi2012}.

While much of the discussion has focused on reproducing the faint \heii\ emission, both of the above scenarios provide a framework to understand two of the most notable features in the UV spectrum of Pox 186: the high EW of \ciii\ and the intense \civ\ emission lines. Multiple studies have used photoionization models to understand the behavior of EW(\ciii) in local and high-$z$ galaxies \citep[e.g.,][]{naka2018,ravi2020}. These models fail to reproduce objects with EW(\ciii) $>$ 20 \AA, which requires intense \ciii\ emission relative to the stellar continuum. While Pox 186 does have intense \ciii\ emission lines, the line intensity ratios are in good agreement with the BOND photoionization models (see Figure \ref{fig:uv_lines}), including those ionized by SEDs with age $\sim$3 Myr (Figure \ref{fig:uv_lines_age}). This suggests that a low stellar continuum is at least partially responsible for the large EW(\ciii), which is consistent with Pox 186's low stellar mass. \citet{ravi2020} and \citet{naka2018} use PopStar and BPASS SEDs that assume a starburst that fully-samples the IMF with a mass of 10$^6$ M$_\odot$, an order of magnitude larger than Pox 186's stellar mass. With a stochastically sampled IMF and low total stellar mass, it may be possible to produce a large EW(\ciii) from intense emission and low stellar continuum (i.e., changing the ratio of ionizing stars to continuum-producing stars). While there are fewer stars that contribute to the UV stellar continuum, there must be massive stars in Pox 186 to produce the highly-ionized gas. As such, an aging stellar population and low stellar mass could simultaneously produce the large EW(\ciii) \citep[and optical \oiii\ EW, see][]{guse2004} and faint \heii\ emission.

As shown in Figure \ref{fig:compspec}, the UV \civ\ emission lines measured in Pox 186 are intense while the \heii\ emission is faint. Both C$^{3+}$ and and He$^{2+}$ exist in the very-high ionization zone, but an amount of C$^{3+}$ is expected in the ionization zone described by O$^{2+}$ emission: photons with energies between 47.8 and 54.4 eV will produce O$^{2+}$ and C$^{3+}$ without producing He$^{2+}$, but photons with energies $\gtrsim$ 55 eV will produce O$^{3+}$, C$^{3+}$, and He$^{2+}$. This energy separation likely leads to the observed difference between the intensities of the UV \civ\ and \heii\ lines. The faint \heii\ emission and lack of observed UV \ion{O}{4} emission lines at 1401.1 \AA\ and 1407.4 \AA\ suggest that very few photons with energy $>$ 54.5 eV are being produced by the ionizing population in Pox 186, but the intense O$^{2+}$ lines in the optical and UV require a significant production of photons with energy $>$ 35.1 eV. In either the aging population or sparsely sampled IMF scenarios, if stars that can produce a small amount of He$^+$ ionizing photons are present then they must produce C$^{2+}$ ionizing photons. The agreement between the \civ\ line ratios (observed and photoionization model) in Figure \ref{fig:uv_lines} indicates that the intense \civ\ is consistent with the emission from other high-ionization ions, which could be produced by an aging stellar population or a small population of massive stars as a result of a stochastically sampled IMF.

If the \heii\ is produced by the ionizing stars, then the true source of the faint \heii\ in Pox 186 and other low-mass SF systems could be a combination of an aging stellar population or a stochastically sampled IMF. Alternatively, the \heii\ is related to other physical mechanisms (such as shocks or HMXBs) that are also sensitive to the age of the ionizing population and sampling of the IMF. As discussed in previous works \citep[e.g.,][]{garn1991,senc2020,oski2022}, these other sources of ionization could produce the discrepancy observed in the optical \heii\ ratios plotted in Figure \ref{fig:optical_lines_age}. Pox 186's intense optical \oiii\ and faint \heii\ emission (both optical and UV) is comparable to the non-extreme GPs and is indicative of recent star formation. Similar to two of the extreme GPs from \citet{jask2019}, the COS UV spectrum reveals absorption from neutral H and low-ionization species of \ion{Si}{2} and \ion{C}{2} despite intense \oiii\ and \ciii\ emission; while this trend could be due to the geometry/covering fraction of the neutral gas along the line of sight, understanding the simultaneous presence of high-ionization metal emission lines and low-ionization metal absorption lines will grant insight into the escape of ionizing radiation through the ISM. If Pox 186 is analogous to the low-mass SF galaxies responsible for reionization, its UV spectrum indicates that these early galaxies may lack very hard ionizing spectra if the IMF is stochastically sampled. 

\section{Conclusions}

We have presented new HST/COS G160M UV spectroscopy of the extreme starburst dwarf galaxy Pox 186. This spectrum is combined with archival COS spectra of Pox 186 from the G130M and G185M filters to give nearly continuous coverage from $\sim$1150 - 2000 \AA. The full UV spectrum reveals the presence of intense, high-EW \oiiisf\ and \ciii\ doublets, double-peaked \civ\ and faint \heii\ emission, and low-ionization absorption features. The EW of the \ciii\ lines is comparable to the the largest in any local starburst galaxy ($>$33 \AA). These lines are used to estimate the density in the highly-ionized gas containing C$^{2+}$, and we find that the density is in the low density limit ($n_e$ = 1$^{+2300}_{-1}$ cm$^{-3}$).

The composite UV spectrum reveals that there is Ly$\alpha$ absorption in Pox 186, in tension with prior \ion{H}{1} 21cm non-detections and indicative of neutral gas along the line of sight to absorb the Ly$\alpha$ photons. This is supported by the low-ionization state absorption features of \ion{Si}{2} and \ion{C}{2}, which have also been observed in high-ionization GPs with Ly$\alpha$ absorption. We model the Galactic and Pox 186 Ly$\alpha$ profiles, then fit the stellar continuum in Pox 186 with a linear combination of \textsc{Starburst99} models to determine the average age and metallicity of the stellar population. We estimate an age and metallicity of $\langle$Age$\rangle = 3.49\pm1.06$ Myr and $\langle$Z$_*\rangle = 0.21\pm0.03 \times$Z$_\odot$ when considering the full range of model ages. When limiting to those models with Age $\leq$ 10 Myr, we determine a similar metallicity but a lower light-weighted age of $\langle$Age$_{<10}\rangle = 2.33\pm0.11$ Myr. The SPS results suggest a recent burst of star formation that has produced predominantly metal-poor stars.

The UV C and O emission lines are used to measure the gas-phase C/O abundance via two methods. The first method directly calculates the C$^{2+}$/O$^{2+}$ ionic abundance from the \ciii\ and \oiiisf\ lines and the optical $T_e$, then corrects this ionic abundance to the total C/O using an ICF. The second method assumes that the C in the ISM can be roughly described by C $\approx$ C$^{2+}$ + C$^{3+}$ and that the total O abundance can be accounted for from the UV \oiiisf\ and optical measurements of the O$^+$/O$^{2+}$ ionic fraction. Both methods produce the same C/O abundance, log(C/O) = $-$0.62$\pm$0.02, in good agreement with other dwarf starburst galaxies of similar metallicity. While the relative C/O abundance is dependent on the star formation history, this agreement suggests that the star formation history of Pox 186 is comparable to other local dwarf galaxies.

Given the utility of \heii$\lambda$1640 in UV emission line diagnostics, we assess the validity of our fit to the faint \heii\ line measured in the new HST/COS spectrum. We employ existing Gemini/GMOS optical IFU spectra of Pox 186 and extract the optical spectrum within a circular aperture of diameter equal to that of COS. While the \oiii\ emission fills the aperture, optical \heii\ emission is compact and located near the center of Pox 186. From the prominent \oiii$\lambda$5007 and narrow \heii$\lambda$4686 lines, we infer the UV \heii/\oiiisf\ line ratio and find good agreement with the intensity ratio measured from the COS spectrum. The \heii\ emission in Pox 186, both from the optical and UV, is significantly fainter than that of local EELGs and extreme GPs. We argue that the observation of faint \heii\ in Pox 186 is possible owing to its proximity and that other highly SF galaxies, like those included in the more general GP sample, may have similarly faint \heii\ emission that is undetected owing to the combination of lower line fluxes and larger noise in the continuum. We also emphasize the potential aperture biases that can be introduced when comparing the emission from very high-ionization (e.g., \heii) and high-ionization ions (e.g., \oiii): for galaxies with extended \oiii\ emission, apertures may exclude significant fraction of the high-ionization zone while simultaneously including most emission from the very-high ionization zone. This effect is exacerbated for local objects that extend beyond the size of most common apertures, potentially producing larger I(\heii)/I(\oiii) ratios in high-ionization systems.

The UV emission lines place important constraints on the ionizing sources in Pox 186. Using the UV emission lines, we produce a series of line ratio diagrams that compare the ratios measured in Pox 186 to those from literature SF galaxies and photoionization models. The log(\civ/\oiiisf) and log(\civ/\ciii) ratios in Pox 186 are in good agreement with the literature sample, consistent with the similar C/O abundance and ionization conditions measured in these sources. Pox 186 is distinct from the literature galaxies when considering the \oiiisf/\heii\ and \ciii/\heii\ ratios, although these ratios agree relatively well with those from photoionization models selected to reproduce giant \hii\ region and BCG optical spectra. We explore two potential explanations that could produce the faint UV \heii/a softer ionizing spectrum in Pox 186: an aging stellar population and a sparsely sampled IMF due to a low total stellar mass. The photoionization models that best reproduce the faint UV \heii\ emission do indicate a stellar population with Age $\geq$ 3 Myr, while other SF galaxies with intense \heii\ can be reproduced by young stars (Age $\leq$ 2 Myr) or stars that have entered the WR phase. However, the photoionization models fail to reproduce the optical \heii/H$\beta$ ratios without invoking higher metallicity models with significant WR phases. This tension could imply that the simple assumptions of single-star, fixed age and metallicity SEDs are inadequate to explain the emission line trends in SF dwarf galaxies, or that other sources of ionization are required to match the intense optical \heii\ emission.

The sparsely sampled IMF is an alternative explanation that is more reasonable for Pox 186 than other local SF galaxies. With such a low stellar mass, the recent star formation in Pox 186 may not have synthesized the high-mass, high-temperature stars necessary to produce a hard ionizing spectrum required for intense \heii\ emission. In such a scenario, the emission line ratios in Pox 186 are not directly comparable to photoionization models which employ SEDs that assume a larger synthesized stellar mass and a fully sampled IMF; dwarf galaxies with significantly larger stellar masses, on the other hand, can more reliably produce these massive stars, resulting in harder ionizing spectra and emission line ratios that are in good agreement with the photoionization models. If the stochastic sampling of the IMF has produced the faint \heii, it has simultaneously produced a highly-ionized environment in Pox 186 that is characterized by intense optical and UV emission from metal ions. Similarly low-mass, reionization-era SF galaxies may be subject to the effects of a stochastically sampled IMF and, therefore, lack the hard ionizing spectrum that can produce a substantial flux of $>$ 54 eV photons.

\begin{acknowledgments}
We thank the referee for their careful review of this manuscript, which has helped refine the discussion and broaden its scope. We thank Simon Gazagnes for sharing his SPS code and for the useful discussion concerning the fitting procedures. We also thank Danielle Berg for her insightful discussions and for sharing her experience using COS observations. This research is based on observations made with the NASA/ESA Hubble Space Telescope obtained from the Space Telescope Science Institute, which is operated by the Association of Universities for Research in Astronomy, Inc., under NASA contract NAS 5–26555. These observations are associated with programs  \#16294, \#16071, and \#16445. Support for program \#16294 was provided by NASA through a grant from the Space Telescope Science Institute, which is operated by the Association of Universities for Research in Astronomy, Inc., under NASA contract NAS 5-03127. J.M.C. is supported by NSF grant AST-2009894. The authors acknowledge Nimisha Kumari for developing their observing programs \#16071, and \#16445.
The Mexican Million Models Database, 3MdB, grid of photoionization models used in the UV emission line analysis is available at \url{https://sites.google.com/site/mexicanmillionmodels}. 


\end{acknowledgments}

\textit{Facilities:} HST (COS), Gemini (GMOS)

\textit{Data:} All the {\it HST}, data used in this paper can be found in MAST: \dataset[10.17909/41cb-0v53]{http://dx.doi.org/10.17909/41cb-0v53}

\textit{Software:} \textsc{AstroPy} \citep{astr2013,astr2018,astr2022}, \textsc{dustmaps} \citep{gree2018}, \textsc{matplotlib} \citep{hunt2007}, \textsc{NumPy} \citep{harr2020}, \textsc{PyNeb} \citep{luri2012,luri2015L}, \textsc{SciPy} \citep{virt2020}, \textsc{SpectRes} \citep{carn2017}.

\bibliographystyle{aasjournal}
\bibliography{pox186_bib.bib}

\end{document}